\documentclass[11pt, reqno,  letterpaper]{amsart}
\usepackage[active]{srcltx}
\usepackage{color}
\usepackage{hhline}
\usepackage{amsmath,amsthm,amssymb}
\usepackage{epsfig}
\usepackage{graphicx}
\usepackage{amssymb}
\usepackage{latexsym}
\usepackage{pdfpages}
\usepackage{bm}
\usepackage{enumitem}
\usepackage{float}
\usepackage{lscape}
\usepackage{braket,booktabs, siunitx}
\usepackage{scalerel}

\usepackage[labelfont=bf, font=small]{caption}

\usepackage{ulem} 

\usepackage{ifpdf}

\ifpdf 
\usepackage[hidelinks]{hyperref}
\else 
\fi 

\usepackage[usenames,dvipsnames]{pstricks}
\usepackage{epsfig}
\usepackage{pst-grad} 
\usepackage{pst-plot} 

\usepackage{color}

\usepackage{subcaption}

\newtheorem*{theorem*}{\it Theorem}

\def\vint_#1{\mathchoice%
	{\mathop{\kern 0.2em\vrule width 0.6em height 0.69678ex depth -0.58065ex
			\kern -0.8em \intop}\nolimits_{\kern -0.4em#1}}%
	{\mathop{\kern 0.1em\vrule width 0.5em height 0.69678ex depth -0.60387ex
			\kern -0.6em \intop}\nolimits_{#1}}%
	{\mathop{\kern 0.1em\vrule width 0.5em height 0.69678ex
			depth -0.60387ex
			\kern -0.6em \intop}\nolimits_{#1}}%
	{\mathop{\kern 0.1em\vrule width 0.5em height 0.69678ex depth -0.60387ex
			\kern -0.6em \intop}\nolimits_{#1}}}
\def\vintslides_#1{\mathchoice%
	{\mathop{\kern 0.1em\vrule width 0.5em height 0.697ex depth -0.581ex
			\kern -0.6em \intop}\nolimits_{\kern -0.4em#1}}%
	{\mathop{\kern 0.1em\vrule width 0.3em height 0.697ex depth -0.604ex
			\kern -0.4em \intop}\nolimits_{#1}}%
	{\mathop{\kern 0.1em\vrule width 0.3em height 0.697ex depth -0.604ex
			\kern -0.4em \intop}\nolimits_{#1}}%
	{\mathop{\kern 0.1em\vrule width 0.3em height 0.697ex depth -0.604ex
			\kern -0.4em \intop}\nolimits_{#1}}}

\def\N{\mathbb N}

\numberwithin{equation}{section}

\def\1{\raisebox{2pt}{\rm{$\chi$}}}

\definecolor{violet(ryb)}{rgb}{0.53, 0.0, 0.69}
\usepackage{mathtools}
\mathtoolsset{showonlyrefs}

\begin{document}
	\title[Chance or Chaos? Fractal geometry aimed to inspect BTC]{Chance or Chaos? Fractal geometry aimed to inspect the nature of Bitcoin}

\author[E. Cabezas-Rivas, F. Sánchez-Coll, I. Tormo]{  Esther Cabezas-Rivas, Felipe Sánchez-Coll and Isaac Tormo Xaixo}

\address{E. Cabezas-Rivas: Departament de Matem\`atiques,
	Universitat de Val\`encia, Dr. Moliner 50, 46100 Burjassot, Spain.
	{\tt esther.cabezas-rivas@uv.es  }}

\address{F. Sánchez-Coll: EDEM Escuela de Empresarios,
Muelle de la Aduana s/n,
46024, Valencia, Spain.
	{\tt fsanchez@edem.es }}

\address{I. Tormo Xaixo: BSc student at EDEM Escuela de Empresarios, Valencia, Spain. 	{\tt istoxa@edem.es}}

\date{\today. The work of the first author is partially supported by the AEI (Spain) and FEDER project PID2019-105019GB-C21, and by the GVA project AICO 2021 21/378.01/1. The first author was also  partially supported by project PID2022-136589NB-I00  as well as the network RED2022-134077-T, both funded by MCIN/AEI/10.13039/501100011033.}

\keywords{Bitcoin, Cryptocurrencies, Fractal geometry, Hurst exponent, Long-term memory, Efficient market.}


	
	\setcounter{tocdepth}{1}
	
	%

	\begin{abstract}
	The aim of this paper is to analyse the Bitcoin in order to  shed some light on its nature and behaviour. We select 9 cryptocurrencies that account for almost 75\% of total market capitalisation and compare their evolution with that of a wide variety of traditional assets: commodities with spot and futures contracts, treasury bonds, stock indices, growth and value stocks.  Fractal geometry will be applied to carry out a careful statistical analysis of the performance of the Bitcoin returns. As a main conclusion, we have detected a high degree of persistence in its prices, which decreases the efficiency but increases its predictability. Moreover, we observe that the underlying technology influences price dynamics, with fully decentralised cryptocurrencies being the only ones to exhibit self-similarity features at any time scale.
	\end{abstract}

	\subjclass[2020]{62P20, 91B82, 91B84.}
	
	\maketitle
	

	\vspace*{-0.5cm}
	
	\section{Introduction}
	
	We face a context of great uncertainty regarding the regulation of Bitcoin (\textsc{btc}). Each country has its own view of its nature, which leads to a wide varie\-ty of legislative solutions ranging from prohibition to full incorporation into economies. However, given the seemingly unlimited opportunities of its underlying technology, global coordination is not only desirable, but necessary.
	
	Indeed, the diversity of approaches creates a confusing landscape for financial institutions and investors seeking to trade \textsc{btc}s. Thus, while it is not surprising that this concern makes headlines in the financial press, it is striking  the multitude of scientific papers with titles as: {\it Is Bitcoin money? And what that means} \cite{Haz}, {\it Is Bitcoin a currency, a technology-based product, or something else?} \cite{White}, or {\it Is Bitcoin a better safe-haven investment than gold and commodities?} \cite{Shaz}.
	
	Paradoxically, we encounter more questions than certainties; therefore, a deep understanding of the behaviour of \textsc{btc}  and similar assets is still  pending, in order for governments to have the knowledge and the right tools to converge towards a standardised regulation.
	
	In  this unsettled framework, the overall goal of this paper is to shed some light on the nature of \textsc{btc} in order to foresee if its prices move randomly or in a chaotic but predictable way, which would allow forecasting opportunities. More precisely, we address the question of how cryptocurrencies (\textsc{cc} hereafter)
	can be classified by means of techniques coming from fractal geometry, and how they
	are correlated with already established asset classes.

We tackle the following key issues:
\begin{enumerate}
\item[{\bf (1)}] Identify factors detecting differences in the dynamics of \textsc{cc} in the markets. \vspace*{-0.2cm}
\item[{\bf (2)}] Understand if there are essential characteristics that make \textsc{btc} peculiar when compared with traditional commodities. The aim is to find out if it can act as diversifying element in portfolio management, or can be exploited as a modern alternative to save haven products.\smallskip
\item[{\bf (3)}] Discern whether the inspection of historical data to predict present and future price evolution makes actual sense, or the moves are aleatory and independent, in which case one better applies probabilistic techniques ad hoc to the concrete distribution.\smallskip
\item[{\bf (4)}] Clarify if return series follow a normal or Gaussian distribution, as most of the methods applied in statistical analysis and prediction models have this as underlying assumption in a direct or subtle way. \smallskip
\item[{\bf (5)}] Judge if it is of any use to take data at various frequencies or subperiods to drive conclusions about prices of \textsc{cc} and other assets, or they present scale invariance that would enable the application of fractal techniques.
\end{enumerate} 

\vspace*{-0.2cm}
	
As highlights of our findings, for all \textsc{cc} with purely decentralized technologies (that is, that really get rid of intermediaries or audit control) we detect clear signs of chaoticity, which do not fit under any traditional Gaussian patterns. Furthermore, the computation of the fractal dimension under different time frames and scales certainly reveals that these \textsc{cc} present long memory and self similarity properties. Hence technical analysis of past data is well-founded but shifting the scale is helpless to get finer predictions. 

We also confirm that these properties are shared  neither by centralized \textsc{cc} nor by  assets like gold, silver, crude, wheat, or treasury bonds, which are quite poorely correlated to \textsc{btc} that is in turn tightly interrelated to most actors of the {\it cryptosphere}. This disables any narrative or marketing strategy to declare \textsc{btc} or as the new gold and its often claimed diversifying power.

Our results upgrade the previous literature in several directions: we pioneer in disclosing that it is the underlying technology, instead of the  capitalization, the crucial element  influencing \textsc{cc} prices; we also open a new research line that challenges the widely accepted impermeability of the \textsc{cc} market to the macroeconomic environment. In fact, we detect a major paradigm shift (driven possibly by rising inflation and interest rates) in \textsc{cc} interrelations and feedback with external products by zooming in on the last trimester of 2022 and 2023 data.

About the techniques we apply to drive the above conclusions, despite they are quite classical going back to Mandelbrot original development of fractal geometry, we enhance them with a proper rigorous mathematical approach by including a priori tests to check that the assumptions of the method are fulfilled by our data, as well as a later inference test  to validate  our outcomes. 

Even though it may seem an obviousness to check the hypotheses before using a procedure and confirm the statistical significance of the computations, as pointed out in \cite{Cou}, many papers in the financial literature ignore naively both steps. Likewise, 
regardless of the more sophisticated statistical tools in the recent literature (as \textsc{dfa} or \textsc{mf-dfa}), as highlighted in \cite{Par}, most of them 
 assume that the variables creating  time series follow a normal distribution, even knowing that this is typically not the case.
 
 To avoid this restriction, we have chosen the robust $R/S$ method \cite{ManTaq, AvTaq} to compute the fractal dimension, since it has minimal requirements, which we ensure that our data accomplish, boosted with  suitable inference tests. We choose this  instead of the modified $R/S$ statistic, as \cite{Ter}  evidenced that the latter produces  strong bias towards
 accepting the null hypothesis of data independence.

On the other hand, there is no need of adding artificial complexity to the computations by using fancy multifractal techniques, as \cite{Bar20} guarantees that most capitalized \textsc{cc} display unifractal patterns, and our study starts with a careful selection of 9 \textsc{cc} representing almost 75\% of the total volume.
	

The paper is organized as follows: in section \ref{back} we introduce the basic notions concerning \textsc{cc} consensus protocols,  memory/efficiency character of time series, and fractality that will be used extensively throughout the paper; whereas section \ref{liter} includes a detailed survey of the literature to stress that the debate about the nature of \textsc{btc} is a vivid area of current research, which has not reached yet any consensus and deserves further attention. Later on, in section \ref{data} we give descriptive statistics of our datasets explaining the sources and criteria of selection of the different assets. Then we explain the details of how we perfom the $R/S$ method (cf.~section \ref{method}) and discuss thoroughly the outcomes and the corresponding implications in section \ref{results}. Finally, we include a summary of main conclusions (section \ref{conclu}) and, for completion, three appendices with additional tables about overview of the techniques used in the previous literature, calculations of the Hurst exponent, as well as full correlation matrices.
	
	\section{Background material: basic facts and definitions} \label{back}
	
	\subsection{Blockchain and its different consensus algorithms}
	
	The paradigm change brought by \textsc{btc} represents an evolution towards decentralised networks, allowing peer-to-peer interactions (with no need of intermediaries or central authorities that audit the operations)  and the secure exchange not only of information but also of value. The technology that enables the latter is known as {\it blockchain}; roughly speaking, it uses cryptographic techniques to run  a  shared and secure digital register to record transactions, like a ledger.

	
		Different consensus protocols/algorithms have been developed to validate and secure operations: {\it  Proof-of-Work} (PoW), where different users ({\it miners}) compete to solve a mathematical problem that requires high computational cost; or	{\it  Proof-of-Stake} (PoS), which selects validators randomly, giving a higher probability to those who deposit a larger amount of \textsc{cc} as a guarantee. 
		
	We will not go into technical details of other variants, like	 {\it Nominated PoS} (NPoS),  {\it PoS Authority} (PoSA),  {\it Delegated PoS} (DPoS), and  {\it Ripple Protocol Consensus Algorithm } (\textsc{rpca}). Let us just point out that some of them are  controlled by very few nodes as in the traditional transaction systems.

	In short,  the consensus method provides each \textsc{cc} with distinct features in terms of energy efficiency, security and scalability; thus, we wonder the following: 
	
	\begin{center}
	\vspace*{-0.1cm} {\it Does the technology of \textsc{cc} influence their behaviour  in the markets?}
	\end{center}
	\vspace*{-0.1cm}
	Accordingly, we will focus on the comparative study of \textsc{cc} with different consensus mechanisms (see Table \ref{pareto-table}). Up to our knowledge, it is the first time that the underlying technology is used in the literature as a distinctive characteristic.
	 
	\subsection{Efficiency versus persistence and fractality}
	
	An essential yet unresolved matter is the predictability of \textsc{cc} behaviour, a notion that challenges the {\it Efficient Market Hypothesis} (\textsc{emh}), which
is one the key cornerstones for modeling financial data \cite{Fama}. A market is efficient if prices  follow a random Brownian motion, reflecting all available information; thus a market is said  efficient if past data cannot be exploited to
predict future returns.  

More precisely,  3 characteristics determine this motion: independence (i.e. prices have no memory and their dynamics are fully random), stationarity (the magnitude of changes does not vary with time) and normality, which implies that extreme events occur with very low probability. The latter cannot  explain the sudden and sharp movements in financial markets. Illustratively, from 1916 to 2003 the Dow Jones index had 48 days with a swing bigger than 7\%, but under the normal distribution  this should occur 1 day every 300,000 years \cite{ManHu}.

In short, the evidence from real data leads to look for  alternatives to \textsc{emh}  that allow to predict these abrupt changes, which also happen in waves and not isolated nor orderly, thus showing the chaotic fluctuation of prices. This is precisely what led B. Mandelbrot to use {\it chaos theory} as an inspiration to create {\it fractal geometry} in the 70s, which seeks to quantify complex patterns  in nature.

Indeed, fractal objects work like chaotic systems where instantaneous shifts can have significant effects  in the long term. Roughly speaking, two main features define a fractal: it has a fractional (non integer) dimension, and it is {\it scale invariant} or {\it self-similar}, i.e.,~presents copies of itself as it is zoomed in (like a snowflake). For a time series this means that its basic features are kept if we consider time subperiods or alter the data frequency of the sample.

When prices evolve inefficiently (then predictably), there are two types of memory: {\it antipersistent} or {\it mean reverting} if an increase is followed by a fall and conversely,  drawing an oscillatory path around the mean; and {\it persistent} or {\it long memory},  that is, after a rise/drop comes another move with the same trend.  In practice, the higher the persistence is, the more difficult it is for the values to return to their predetermined target in the event of a fall or exogenous shock.


	\vspace*{-0.3cm}
	
	\section{Literature review} \label{liter}
	
	\subsection{Under which label do we classify Bitcoin?}
	
	The right description of \textsc{btc} is a desirable goal  to grasp its potential role in the market for risk management and portfolio diversification. While its design has similarities with gold (mining, decentralization, not government-backed, globally traded 24/7) and currencies (medium of exchange),  if \textsc{btc} were a real unit of account or a store of value, it would not exhibit high volatility
	characterized by bubbles and crashes (cf.~\cite{ChFr}). 
	
	In this spirit, it resembles more a highly speculative asset than a typical commodity or currency (see \cite{Gaj}), or at least belongs to a category in between the latter two (cf.~\cite{Dyh}). Moreover, \cite{Baur} shows that \textsc{btc} has its own risk-return features and is
	uncorrelated with traditional assets.  Because of this, \cite{Klein} concludes that it cannot play the role of a safe-haven from an
	econometric perspective.
	
	In fact, its evolution does not respond to monetary policy news, but it reacts to  events related to \textsc{cc}, having significant correlations with them  \cite{assaf, Vidal}. 	
	Ana\-logously, \cite{Ji} claims that the
  isolation of \textsc{btc} from the global financial system implies that it is not an actual source of economic instability.
	
However, the  dependence  between \textsc{cc} and
	other assets may change over time. Indeed,  \cite{ZM} checks that the gold-\textsc{btc} correlation reached a maximum during the peak of the \textsc{covid}-19, dropping  to almost zero in July 2021. Because of this, 
	 it is not completely hopeless to include \textsc{cc} in a portfolio. Even more, \cite{Dyh} argues that \textsc{btc}  is helpful for risk-averse investors
in anticipation of negative shocks, as its reactions to market sentiment are quicker.

	\subsection{Efficient Market Hypothesis (EMH) in the \lq\lq cryptosphere"}

Des\-pite the extensive analysis about  \textsc{emh}'s applicability to the \textsc{btc} market  (see \cite{Corb} and Table \ref{lit:table} for a
 survey),  there is no agreement on whether the periods of efficiency alternate either with mean reverting dynamics (as claimed in \cite{Alv, Urq}) or with long memory trends (stated by \cite{Bar1,Phi}).

 More precisely, \cite{Capo} unveils a decreasing trend in the predictability, confirmed in \cite{Koch} by checking a reduction in price reaction time to unexpected events. This  is further supported in \cite{Urq2}, as no pattern in returns can be discovered away from price clustering. But there is no common narrative to explain the varying efficiency over time:  according to \cite{Kris},  inefficiency is higher during price rises;  \cite{Sens} found out that liquidity (volatility) has a significant positive (negative) effect on the 
 efficiency, which can be enhanced by introducing \textsc{btc} futures \cite{Ruan}.
 
  On the contrary, there are sources claiming that \textsc{btc} moves similarly most of the time, but there is no agreement between those who advocate that the mainstream is efficiency \cite{Tiw, Nada},  long term memory \cite{AlY, Hu, jiang}, or   anti-persistency \cite{Sta}, no matter the frequency or time frame considered.

Despite the almost exclusive restriction to \textsc{btc}, some authors  compare it with other \textsc{cc}. In this spirit,  \cite{Braun} concludes that \textsc{btc} is the least predictable, which  is reinforced by \cite{Bar20}, with notable efficiency in lower volume quantiles and anti-persistance in higher ones. However, \cite{Dav, Men} also find evidences of the reverse arguments, showing that less capitalized coins are
more efficient than  \textsc{btc}, while the latter presents long memory (see also \cite{Chaf}). 

Finally, efficiency can be altered during exceptional circumstances, such as the \textsc{covid}-19 pandemic, which introduced significant regime changes in crypto and traditional markets (see \cite{Lah3}). Surprisingly enough,   \textsc{btc} is more resilient to efficiency decreases than other financial assets (cf. \cite{WW}).

	\subsection{Unraveling complexity by means of fractal geometry}

		\cite{Aslan, Oma} claim that the efficiency  of \textsc{cc} varies across frequencies, being this heterogeneous memory behaviour against the self-similarity required for fractal objects. This view is shared by 
		\cite{Sens}, which checked that higher the frequency, lower the pricing efficiency is. In the same spirit, 	\cite{Lah1, Men} conclude that the regime of persistence depends on the time scale and the period considered. 
		
		On the contrary, \cite{Bar2} reports similar memory patterns, no matter of the time frequencies,
		implying a self-similar process. This is confirmed in \cite{Lah2} via a big data-driven study joint with statistical testing, providing evidence of dominant fractal traits at all high frequency rates for \textsc{btc} prices.
		
			Later on, \cite{Bar20}   clearly reveals that one cannot apply a common model to address the whole \textsc{cc} landscape. Indeed,  highly capitalized  coins display roughly unifractal processes, and thus can be described via fractional Brownian
		motion; but cryptoassets with fewer liquidity  exhibit strong multifractality, and more sophisticated models would be required for capturing their complex dynamics.

	As pointed out in \cite{Sto}, multifractal features are not exclusive of \textsc{cc}, in fact, they are similar to that of stock markets, but differ from
	 regular coins. Additionally, \cite{tell} checked that \textsc{btc} shows  higher fractality
than gold,  which reopens the debate about its nature also from the outlook of fractal geometry.

	\section{Data and descriptive statistics} \label{data}
	
	\subsection{Types of assets: selection criteria and sources}

	 We have worked with a dataset (obtained from {\it Binance}) containing the prices of \textsc{btc} and eight other \textsc{cc}  (see Table \ref{description-all}) over the time period from 20/8/2020 to 24/2/2023, considering opening values every 15 minutes ($N = 88,\!060$ observations), 1 hour ($N = 22,\!019$) and 1 day ($N = 919$). The aim is to contrast daily dynamics with intraday or high frequency movements in view of seeking for self-similarity features.

	In order to compare with other assets traded on traditional financial markets, time series have been selected from {\it Bloomberg} with daily values of:
	\begin{itemize}
		\vspace*{-0.3cm}
		\item[$\circ$] 3 commodities with futures contracts (gold, crude oil and wheat), 
		\item[$\circ$] one spot commodity (silver),
		\item[$\circ$] a 10-year US Treasury bond futures contract (FC), 
		\item[$\circ$] 2 stock market indices (Nasdaq and EuroStoxx), 
		\item[$\circ$] 3 {\it growth} stocks (Tesla, Netflix and Amazon), and  3 {\it value} stocks (The Coca-Cola Co., Procter \& Gamble and Johnson \& Johnson).
	\end{itemize}

	\renewcommand{\arraystretch}{1.3} 
\begin{table}[H] 
	\fontsize{7pt}{7pt}\selectfont
	\centering
	\resizebox{\textwidth}{!}{\begin{tabular}[H]{lccl}
		\toprule Asset & Ticker  & Market & Initial Release/Public Offering/Description  \\
		\midrule \midrule
		Bitcoin & BTC & Binance & 9 January 2009 \\
		Ethereum & ETH  & Binance & 30 July 2015 \\
		Binance Coin & BNB  & Binance &  3 July 2017 \\
		Ripple & $\mathrm{XRP}$  & Binance & June 2012 \\
		Cardano & $\mathrm{ADA}$  & Binance & 	27 September 2017  \\
		Polygon & MATIC  & Binance & 2019 \\
		Solana & SOL  & Binance & 24 March 2020 \\
		Tron & TRX  & Binance & 25 July 2018 \\
		Polkadot & DOT & Binance & 26 May 2020 \\
		Amazon & AMZN & NASDAQ & 15 May 1997 \\
		Tesla & TSLA  & NASDAQ & 29 June 2010 \\
		Netflix & NFLX  & NASDAQ & 23 May 2002 \\
		Procter \& Gamble & PG  & NYSE & 13 Jan 1978 \\
		Johnson \& Johnson & JNJ  & NYSE & 24 Sept 1944 \\
		The Coca-Cola Co. & KO  & NYSE & Sept 1919 \\
		Silver & XAGUSD  & LBMA & spot price per ounce in US dollars \\
		Gold & GCG23  & COMEX & 100 troy ounce FC due in Feb 2023\\
		Crude oil & CLK23  & NYMEX & West Texas Intermediate  FC due May 2023 \\ 
		Wheat & ZWK23 & NYMEX & FC expiring in May 2023 \\
		US Treasury bonds & ZNM23  & $\mathrm{CBOT}$ &  10-year FC due in June 2023\\
		Nasdaq & CCMP  & NASDAQ & launched in 1971 \\
		EuroStoxx & SX5E  & EUROSTOXX & 26 February 1998 \\
		\bottomrule
	\end{tabular}} \caption{Description of cryptoassets and traditional commodities} \label{description-all}
\end{table}

 \vspace*{-0.3cm}
 We  conjecture  that growth stocks (i.e., those with a 5-year average sales growth over 15 \%) will perform similar to cryptoassets having actually positive correlation; while it is expected that when \textsc{cc}  go up, value stocks (that is, the ones with price-to-sales ratio $< 1$) will go down.

 On the other hand, to assess whether \textsc{cc}  perform as a safe haven asset, commodities have been chosen, as well as low volatility assets as treasury bonds. Moreover, to compare the cryptoeconomy to the traditional financial market we include stock indices, which act as a thermometer of market movements.

Regarding the selection of \textsc{cc}, we applied the Pareto principle to their capitalisation, according to which 80\% of the results are due to 20\% of the  variables involved. Although the 9 selected \textsc{cc} \lq\lq only" accumulate approximately 74\% of the total market capitalisation (see Table \ref{pareto-table}), we have ruled out \textsc{cc} with volume percentages below 0.5\%, which we consider to be unrepresentative. Let us stress that we have chosen archetypes of five different consensus protocols to enable our aim of comparing \textsc{cc} on the basis of their underlying technology.

\renewcommand{\arraystretch}{1.3} 
\begin{table}[H]
	\fontsize{7pt}{7pt}\selectfont
	\centering
\begin{tabular}{llllll}
		\toprule
		Ticker & Protocol & Capitalization (million \$) & Percentage & Cumulative & \textbf{} \\ \midrule \midrule
		BTC & PoW & 517,890 & 45.51\% & 45.51\% & ~ \\ 
		ETH & PoS & 218,230 & 19.18\% & 64.69\% & ~ \\ 
		BNB & PoSA & 48,060 & 4.22\% & 68.91\% & ~ \\ 
		XRP & RPCA & 23,590 & 2.07\% & 70.98\% & ~ \\ 
		ADA & PoS & 12,760 & 1.12\% & 72.10\% & ~ \\ 
		MATIC & PoS & 10,080 & 0.89\% & 72.99\% & ~ \\ 
		SOL & PoS & 7,690 & 0.68\% & 73.66\% & ~ \\ 
		TRX & DPoS & 6,960 & 0.61\% & 74.28\% & ~ \\ 
		DOT & Nominated PoS & 6,300 & 0.55\% & 74.83\% \\ \bottomrule
	\end{tabular} \caption{Description of the 9 selected \textsc{cc} (from  \url{www.binance.com}, 24 May 2023)} \label{pareto-table}
\end{table}

\begin{figure}[H] 
	\centering
	\includegraphics[scale=0.35]{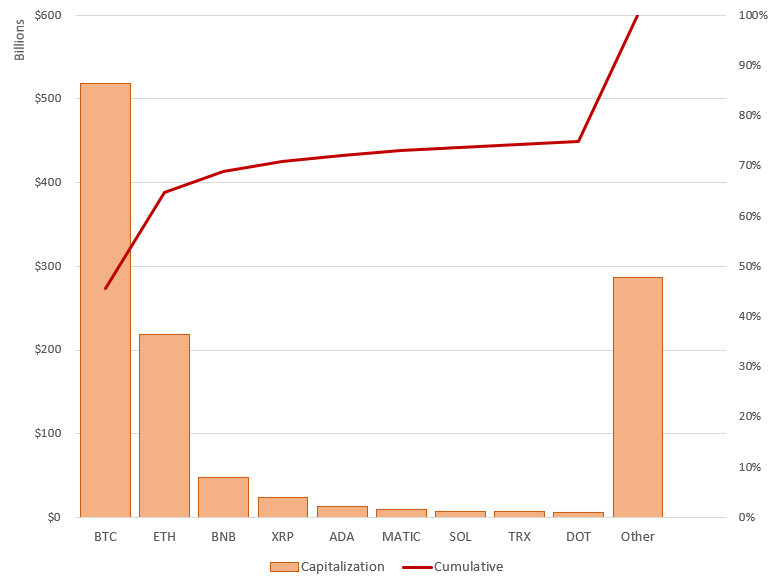}  \caption{Pareto diagram of the cryptocurrency market according to capitalization} \label{pareto:fig}
\end{figure}

\vspace*{-0.3cm}
As can be seen in Figure \ref{pareto:fig}, \textsc{btc} and \textsc{eth} account for almost 65 \% of the total volume, while the 9 chosen \textsc{cc} together have a total capitalisation of more than \$851 billion, which will give us a global view of the market, accounting for almost 75\% of the total capitalisation estimated at \$1,138 trillion, according to Investing (\url{www.investing.com}) as of 24 May 2023.

\vspace*{-0.3cm}
\subsection{Descriptive statistical analysis of data}

Let us first point out that daily opening values have been taken. In addition, unless otherwise specified, to facilitate the direct comparison, only weekday values are included, even though \textsc{cc} are still trading at weekends.

To analyse the normality of the time series, one typically converts them  from price series $P=\{P_t \, : \,  t\in \N\}$ to return series $R=\{R_t \, : \, t\in \N\}$. In our case, we have calculated a total of 608 returns for each series on a logarithmic scale, i.e. we work with the sample $R=\{R_t \, : \, t=1,\ldots,608\}$, where
\begin{equation} \label{return}
R_t:= \log \frac{P_t}{P_{t-1}}=\log P_t-\log P_{t-1}.
\end{equation}

\vspace*{-0.3cm}
Table \ref{descriptive} reports the descriptive statistics for the full sample period from 2/8/2020 to 24/2/2023, and leads us to the following preliminary conclusions:

\begin{itemize}
	\vspace*{-0.3cm}
	\item[$\circ$] \textsc{cc} have a much higher volatility than other assets; in particular, \textsc{btc}'s volatility (3.85) is much higher than gold's (0.92). As a result, this is a first indication that \textsc{btc} does not behave as a safe haven asset.
	
	\item[$\circ$] For a normal distribution, the kurtosis  
	is around 3 and the skewness is near zero. The values of the latter do not seem conclusive, as most of them are close to zero; however, the kurtosis seems to point that neither \textsc{cc} nor traditional stock market players follow a normal pattern. 
	
	\item[$\circ$]  The normality test (Jarque Bera) brings strongly significant evidence that none of the series under study follows Gaussian moves, which confirms Mandelbrot's criticism \cite{ManHu} to classical economic theories arguing that the normality assumption does not properly capture price evolution.
\end{itemize}

\renewcommand{\arraystretch}{1.2} 
\begin{table}[H]
\fontsize{7pt}{7pt}\selectfont
\centering
	\resizebox{\textwidth}{!}{	\begin{tabular}[H]{lrrrrrrrr} 
 \toprule  Asset & Mean & Median & Std. & \multicolumn{1}{c}{ Max. } & \multicolumn{1}{c}{ Min. } & Skew. & Kurt. & J. Bera\\
		\midrule \midrule   BTC & 0.0401 & -0.0190 & 3.8451 & 17.8448 & -16.7093 & -0.2419 & 5.9025 & 376.16$^{***}$\\
		ETH & 0.1428 & 0.2390 & 5.1954 & 23.3707 & -32.4864 & -0.4250 & 7.5343 & 288.07$^{***}$\\
		BNB & 0.2217 & 0.1630 & 5.6390 & 29.5648 & -41.6751 & -0.2158 & 11.5924 & 7311.6$^{***}$ \\
		XRP & -0.1597 & 0.1424 & 6.6802 & 36.6201 & -53.8523 & -0.6736 & 15.8684 & 3289.8$^{***}$\\
		ADA & -0.0909 & -0.1909 & 5.9398 & 28.7239 & -31.1317 & 0.1538 & 6.5777 & 435.14$^{***}$\\
		\textsc{matic} & 0.4301 & -0.0706 & 8.1566 & 48.7557 & -41.0080 & 1.0209 & 9.6473 & 1639.1$^{***}$\\
		SOL & 0.0081 & -0.2235 & 7.9623 & 38.0494 & -54.9008 & -0.5784 & 9.8461 & 665.94$^{***}$\\
		TRX & 0.0556 & 0.1906 & 5.5029 & 34.3355 & -38.8245 & -0.2373 & 11.5803 & 1096.2$^{***}$\\
		DOT & -0.0957 & -0.2126 & 6.5311 & 28.0615 & -48.3208 & -0.3569 & 10.0600 & 271.22$^{***}$\\
		AMZN & -0.1167 & 0.0016 & 2.5189 & 10.4044 & -15.1499 & -0.3680 & 6.4681 & 296.59$^{***}$\\
		TSLA & 0.0707 & 0.1384 & 4.3051 & 14.4446 & -17.0308 & -0.1375 & 4.3593 & 47.421$^{***}$\\
		NFLX & -0.1004 & -0.0627 & 3.0387 & 12.0963 & -30.6729 & -2.4777 & 28.1416 & 15228$^{***}$\\
		PG & 0.0034 & 0.0768 & 1.1469 & 3.6368 & -7.5586 & -0.8748 & 7.1694 &  498.61$^{***}$\\
		JNJ & -0.0043 & 0.0270 & 1.0430 & 5.5584 & -3.7804 & 0.1567 & 4.6431 & 66.278$^{***}$\\
		KO & 0.0419 & 0.0869 & 1.1405 & 5.5815 & -7.0610 & -0.2622 & 6.9038 &  333.44$^{***}$\\
		Silver & -0.0544 & -0.0637 & 1.8350 & 7.9983 & -8.5103 & -0.1167 & 5.6815 & 154.02$^{***}$\\
		Gold & -0.0143 & 0.0429 & 0.9210 & 2.8140 & -4.8252 & -0.5177 & 4.9956 & 134.18$^{***}$ \\
		Crude & 0.0892 & 0.2208 & 2.5703 & 11.6753 & -12.3624 & -0.1827 & 5.1919 & 181.26$^{***}$\\
		Wheat & 0.0630 & -0.1375 & 2.2508 & 17.5554 & -9.6304 & 0.8261 & 9.6193 & 1179.1$^{***}$ \\
		Bonds & -0.0303 & -0.0036 & 0.4281 & 1.7628 & -1.4895 & 0.0768 & 4.3337 & 46.587$^{***}$\\
		Nasdaq & -0.0005 & 0.1616 & 1.6119 & 6.8863 & -7.0825 & -0.3765 & 4.4745 & 22.102$^{***}$\\
		EuroSt. & 0.0310 & 0.1008 & 1.0992 & 5.6815 & -5.3614 & -0.2227 & 5.9040 & 276.7$^{***}$\\
		\bottomrule 
	\end{tabular}} \caption{Descriptive statistics and normality test (Jarque Bera) for daily returns from  20/8/2020 to 24/2/2023. $^{***}$ means significant results at 0.1\% level of significance.} \label{descriptive}
\end{table}

\vspace*{-0.3cm}
Additionally, we split our sample into two
subperiods (see Table \ref{subperiods}) in order to test whether the  efficiency has varied
over time. 
A first inspection shows that descriptive statistics are quite stable for \textsc{btc}, while other assets present sign shifts in mean (\textsc{trx, nflx}) or kurtosis (silver). This provides a first hint that self-similarity features may not be shared for all items under review. Let us also point out that the significativity level for the normality test raises to 5\% during the second subperiod for wheat, oil, gold, silver, Nasdaq, US bonds, \textsc{tsla} and \textsc{jnj}, which may foresee a change in performance.

\renewcommand{\arraystretch}{1.2} 
\begin{table}[H]
	\fontsize{7pt}{7pt}\selectfont
	\centering
	\begin{tabular}{llllll}
		\toprule Sample period & N & Mean & SD & Skew. & Kurt. \\
		\midrule	\midrule   {\bf BTC} & & & & \\
		20/8/2020 -- 01/7/2022 & 681 & 0.0395 & 4.1134 & -0.1986 & 5.2739  \\
		01/7/2022 -- 24/2/2023 & 239  & 0.0358 & 2.9172 & -0.5396 &  8.9691 \smallskip \\
		{\bf TRX} & & \\
		20/8/2020 -- 01/7/2022 & 681 & 0.1229 & 6.2265 & -0.2374 & 9.4431 \\
		01/7/2022 -- 24/2/2023 &239  & -0.1431 & -0.6032 & -0.5396 &  5.9418  \smallskip \\
		{\bf NFLX} & & \\
		20/8/2020 -- 01/7/2022 &470 & -0.2597 & 3.0527 & -3.4819 & 34.9563 \\
		01/7/2022 -- 24/2/2023 & 164 & 0.36498 & 2.9482 & 0.7096 &  3.9657
		\smallskip \\
		{\bf Silver} & & \\
		20/8/2020 -- 01/7/2022& 485 & -0.0710 & 1.7950 & -0.4791 & 6.0914 \\
		01/7/2022 -- 24/2/2023 & 169 & -0.0199 & 0.7124 & 0.7096 &  4.5917\\
		\bottomrule
	\end{tabular}
	\caption[width=\textwidth]{Extract from the descriptive statistics of the two subsample periods.} \label{subperiods}
\end{table}

\vspace*{-0.3cm}

Finally, let us mention that we do not display descriptive statistics for intraday frequencies of \textsc{cc} for brevity, since there are not remarkable differences with Table \ref{descriptive}, apart from the higher number of observations.

\section{Methodology: R/S analysis enhanced by a test of significance} \label{method}

A fractal is defined by scale invariance and chaotic nature, whose complexity is measured by the fractal dimension $D$. As the graphs of returns are bumpy/peaky curves in the plane, $D$ should be between 1 (dim.~of a smooth curve) and 2 (dim.~of the plane); then, we can write $D=2\ -H$, with $0 < H <1$. 

For time series one typically computes the value of $H$, known as {\it Hurst exponent}, which was introduced to study the Nile overflows \cite{hurst1, hurst2}. Economists use $H$ to assess the efficiency of the market, so that it is considered efficient if $H=0.5$, and inefficient otherwise (see more details in Table \ref{Hurst-int}).

	\renewcommand{\arraystretch}{1.2} 
\begin{table}[H] 
	\fontsize{7pt}{7pt}\selectfont
	\centering
	\resizebox{\textwidth}{!}{	\begin{tabular}[H]{lccc}
			\toprule  &  $H = 0.5$  & $0 < H < 0.5$ & $0.5 < H < 1$  \\
			\midrule \midrule
			Motion & random Brownian & fractional Brownian & fractional Brownian \\
			Persistence & none (independent)  & anti-persistent (mean-reverting) & persistent \\
			Memory/Correlation & none  & short term &  long term \\
			Efficiency & efficient  & inefficient & inefficient \\
			\bottomrule
	\end{tabular}} \vspace*{-0.2cm} \caption{Interpretation of the values of the Hurst exponent} \label{Hurst-int}
\end{table}

\vspace*{-0.3cm}
Hereafter we explain the rescaled range ($R/S$) method (cf.~\cite{Man1, Man2, ManTaq}) to get $H$ without imposing independence nor  normality of the returns (as other more methods widely used in the literature do). The only restriction is to work with stationary series, but we will check that this is the case (see subsection \ref{stat:sec}).

One of the additional advantages of the $R/S$
analysis is that it is robust in the sense that can detect non-periodic cycles even if they are longer than the 
sample period, as well as long-term correlations. For instance, \cite{ManWa} applied it to conclude that many natural phenomena are not independent random processes.

\begin{figure}[H] 
	\centering
	\includegraphics[scale=0.5]{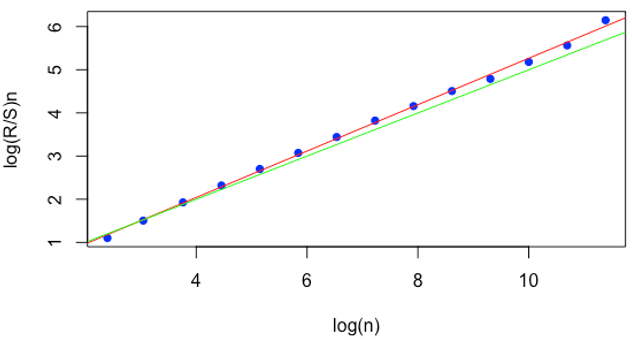}  \vspace*{-0.3cm}\caption{Graphic representation of the Hurst exponent. The slope of the red line represents the Hurst exponent while the green line has a slope of 0.5.} \label{regression:fig}
\end{figure}

We have divided the calculation procedure into the steps described below:

\begin{enumerate}
	\item[{\bf (1)}] Take  a series of returns $R=\left\{R_1,\cdots,R_N\right\}$, computed as in \eqref{return}. 
	
	\vspace*{0.2cm} 
	
	\item[{\bf (2)}] Split the full series $R$ into $d$ sub-series of length $n=\frac{N}{d}$:
		\vspace*{0.1cm} 
	\begin{itemize}
	\item[$\circ$] 1st sub-series: $\{R_1^1,\cdots,\ R_n^1\}$, 	\vspace*{0.1cm} 
	\item[$\circ$] 2nd sub-series: 	\vspace*{0.1cm}  $\{R_1^2,\cdots,\ R_n^2\}$, and so on up to the 
	\item[$\circ$] $d^{\,\text{th}}$ sub-series: $\{R_1^d,\cdots, R_n^d\}$. 	\vspace*{0.1cm} 
	\end{itemize}
	In short, we consider  a pack $\displaystyle \{R_i^m\}_{{i=1,\ldots,n; m=1,\ldots,d}}$ with all sub-series. This step is done with $d=1,2,3,\ldots,\frac{N}{n}$ for different values of $6<n<N$.
	
		\vspace*{0.2cm} 
	\item[{\bf (3)}] For each sub-series $R_i^m$,  compute the mean $E_m$ and the standard deviation $S_m$ (which will be functions of $n$).
	
		\vspace*{0.2cm} 
	\item[{\bf (4)}] Determine the series of distances to the mean $Z_i^m$ by means of 
\[	Z_i^m=R_i^m-E_m,\]
	and create the cumulative time series $Y_i^m$ for each sub-series $m=1,2,\ldots,d$:
	\[Y_i^m=Z_1^m+Z_2^m+\cdots Z_i^m=\sum_{j=1}^{i}Z_j^m.\]

	\vspace*{0.2cm} 
	\item[{\bf (5)}] Find the range $\bm{R}_m$ of each cumulative subset, for all $m$:
	\[\bm{R}_m(n)=\max\{Y_1^m,Y_2^m,\ldots,Y_n^m\}-\min\{Y_1^m,Y_2^m,\ldots,Y_n^m\}.\]
	Then divide this by the corresponding standard deviation $S_m$, that is, compute $\bm{R}_m/S_m$. Thus, one gets a dimensionless measure  that depends on $n$ and allows comparing the relative variability of sets of different sizes.
	
		\vspace*{0.2cm} 
	\item[{\bf (6)}] Obtain the rescaled range statistic ${(\bm{R}/S)}_n$ by averaging for all sub-series:
	\[{(\bm{R}/S)}_n=\frac{1}{d}\sum_{m=1}^{d}\frac{\bm{R}_m}{\ S_m}(n)\]
	Note that the different partition sizes ($d$) from {\bf (2)} lead to the set of values:
	\[\Big\{{(\bm{R}/S)}_N,\ {(\bm{R}/S)}_\frac{N}{2},\ {(\bm{R}/S)}_\frac{N}{4},\ \cdots,\ {(\bm{R}/S)}_6\Big\}\ =\ \ \left\{{(\bm{R}/S)}_n\right\}_{n=N, \frac{N}{2}, \frac{N}{3},\ldots}\]
	
		\vspace*{0.2cm} 
		\item[{\bf (7)}] {\bf Computation of $\bm{H}$}.  By analogy to Hurst's ideas, assume that the  varia\-bility of the data follows a potential law of the form:
	\[{(\bm{R}/S)}_n = c \cdot n^H,\]
	now take logarithms on both sides to reach the linear relation 
	\[\log(\bm{R}/S)_n=\log c+ H\log n.\]
	We can plot $(n, \log(\bm{R}/S)_n)$ for $n \in \left\{N,\frac{N}{2},\frac{N}{3},\ldots\right\}$, and get the corresponding regression line (see Figure \ref{regression:fig}), whose slope is the desired $H$.
\end{enumerate}
\vspace*{-0.3cm}

\begin{figure}[H] 
	\hspace*{-0.2cm}	\includegraphics[scale=0.38]{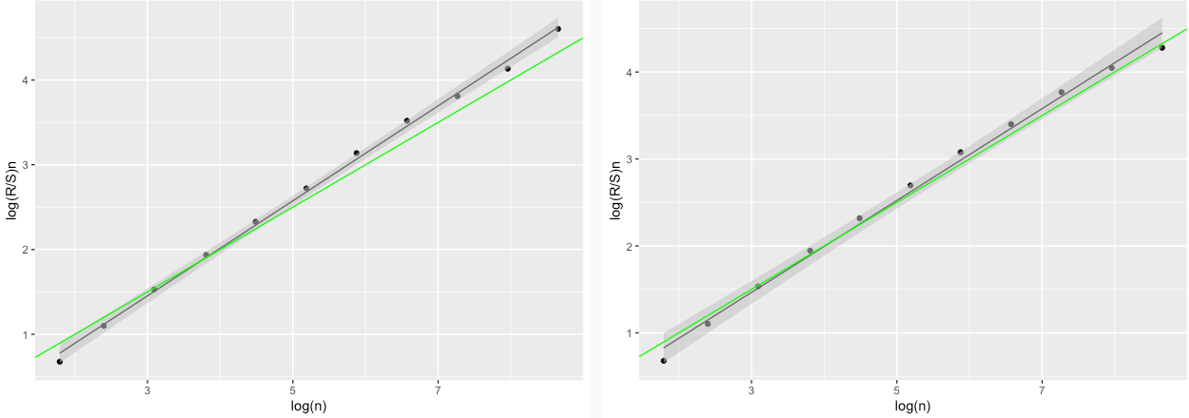}  \vspace*{-0.2cm}\caption{Graphical example of estimation of Husrt exponent (slope of the black line) and its 99\% confidence interval (shadow region) for a long memory time series (left) and a random process (right). The green lines have slope exactly 0.5.} \label{confidence:fig} 
\end{figure}

\vspace*{-0.3cm}
Many papers in the literature stop here, which is a naive approach as pointed out in \cite{Cou}, since one needs an extra  statistical  test to judge if  $H$ computed for the sample is significatively different from the value 0.5 characterizing an
independent process. This is why we have developed our own  R script to top up the $R/S$ method by adding a $t$-test for the slope of the line (see Figure \ref{confidence:fig}).

\section{Results and discussion} \label{results}

\vspace*{-0.3cm}

\subsection{Test for stationarity as prerequisite for R/S analysis} \label{stat:sec}

Despite this step is often skipped, to be rigorous, one needs to check that the assumptions of a method are fulfilled before applying it, in this case, that all time series are stationary. For the \textsc{cc} the corresponding test ensures at more than 99\% confidence that this is the case (see Table \ref{station-cc}) at any scale in the full interval (these results are not altered for the subperiods).
\renewcommand{\arraystretch}{1.2} 
\begin{table}[H]
	\fontsize{7pt}{7pt}\selectfont
	\centering
	\begin{tabular}{lccc}
		\toprule
		& 15 min (lag order = 44) & 1 h (lag = 28) &  daily (lag = 9) \\ \midrule \midrule
		ADA & -44.340$^{**}$ & -29.644$^{**}$ & -8.2618$^{**}$ \\ 
		BNB & -42.009$^{**}$ & -29.384$^{**}$ & -7.5603$^{**}$ \\ 
		BTC & -43.426$^{**}$ & -28.566$^{**}$ & -8.7709$^{**}$ \\ 
		DOT & -43.915$^{**}$ & -29.196$^{**}$ & -9.8369$^{**}$ \\ 
		ETH & -43.757$^{**}$ & -28.549$^{**}$ & -9.1639$^{**}$ \\ 
		MATIC & -43.159$^{**}$ & -29.456$^{**}$ & -9.3751$^{**}$ \\ 
		SOL & -44.326$^{**}$ & -29.367$^{**}$ & -8.3510$^{**}$ \\ 
		TRX & -43.360$^{**}$ & -30.270$^{**}$ & -9.0032$^{**}$ \\ 
		XRP & -43.736$^{**}$ & -27.587$^{**}$ & -8.9680$^{**}$ \\ \bottomrule
	\end{tabular}
\vspace*{-0.2cm} \caption{Results of the Augmented Dickey-Fuller (ADF) test for stationarity for cryptoassets with different frequencies  for the period 20/8/2020 -- 24/2/2023. $^{**}$ concludes that all series are stationary at less than 1\% level of significance. } \label{station-cc}
\end{table}

\vspace*{-0.3cm}	
Concerning the remaining assets, stationarity is guaranteed at the same level of significance for daily data in all the three periods under study, as can be checked in Table \ref{station-other}.
\renewcommand{\arraystretch}{1.2} 
\begin{table}[H]
	\fontsize{7pt}{7pt}\selectfont
	\centering
	\begin{tabular}{lccc}
		\toprule
		& full period (lag order = 8) & 1st subset (lag = 7) & 2nd subset (lag = 5) \\ \midrule \midrule
		Gold & -8.7680$^{**}$   & -8.7733$^{**}$ & -5.2928$^{**}$ \\ 
		Silver & -8.8266$^{**}$   & -8.1462$^{**}$ & -5.6915$^{**}$ \\ 
		Nasdaq & -8.0335$^{**}$   & -8.5723$^{**}$ & -5.1938$^{**}$ \\ 
		Eurotock & -7.8613$^{**}$   & -8.1981$^{**}$ & -5.1350$^{**}$ \\ 
		Oil & -10.0500$^{**}$   & -9.6601$^{**}$ & -6.7996$^{**}$ \\ 
		US bonds & -8.1775$^{**}$   & -7.6914$^{**}$ & -4.9403$^{**}$ \\ 
		Wheat & -6.8168$^{**}$  & -5.5172$^{**}$ & -4.8443$^{**}$ \\ 
		AMZN & -8.1451$^{**}$   & -8.2698$^{**}$ & -5.1547$^{**}$ \\ 
		NFLX & -7.8880$^{**}$   & -7.6475$^{**}$ & -5.2018$^{**}$ 	\\ \bottomrule
	\end{tabular}
\vspace*{-0.2cm}\caption{Results of the Augmented Dickey-Fuller (ADF) test for stationarity for traditional assets with daily values for the three different periods. $^{**}$ concludes that all series are stationary at less than 1\% level of significance. } \label{station-other}
\end{table}

\begin{figure}[H] 
	\centering
\hspace*{-0.3cm}	\includegraphics[scale=0.39]{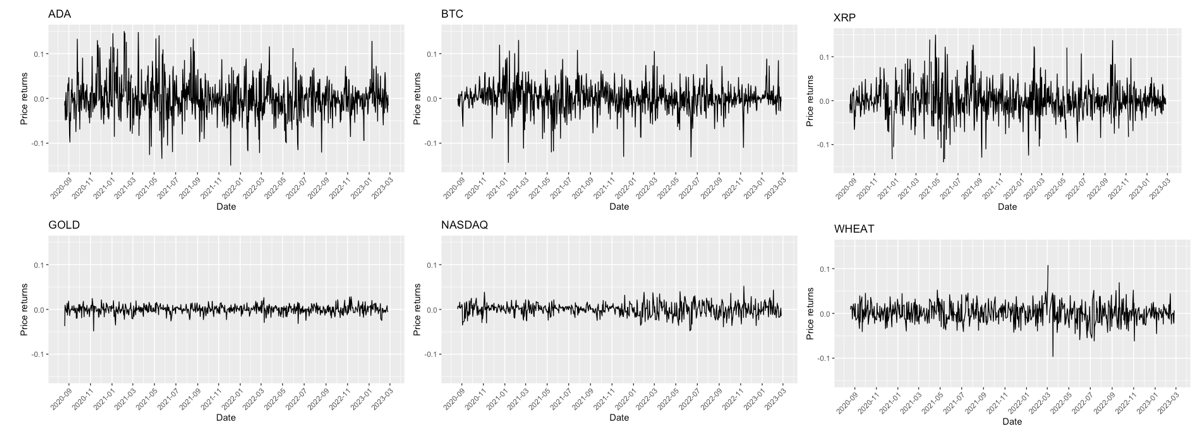}  \vspace*{-0.4cm} \caption{Comparison of daily return graphs for several assets during the whole period. } \label{returns-fig}
\end{figure}

\vspace*{-0.3cm}
Let us stress that, regardless the stationary character of all the return series, a quick visual inspection (see Figure \ref{returns-fig}) already gives us an early warning about the different  complexity level in the graphs of \textsc{cc} versus traditional assets (even through the curves for \textsc{cc} are drawn with average daily values to smooth out irregularities). The tool to quantify this and grasp finer disparities within \textsc{cc}, which are not displayed at first sight by the graphs, is the fractal dimension or, equivalently, the Hurst exponent that we compute in the sequel.

\vspace*{-0.5cm}

\subsection{Memory and efficiency in cryptocurrencies and traditional assets}

We start by analysing the Hurst index (see Table \ref{Hurst-1}) of the time series for the full temporal frame with daily opening data for \textsc{cc}, commodities, stock indices and the US bonds. At a first glance, we conclude that all \textsc{cc} (except \textsc{xrp}) have long-term memory, and thus it would make sense to carry out a technical analysis to predict their future behaviour. As a corollary, \textsc{cc} do not fulfil the \textsc{emh} of statistical independence of prices. 

The same is true for most of the remaining assets, but our data do not provide enough evidence to discard that silver or gold  move randomly; hence, they \lq\lq forget" the past.  Nasdaq and two of the stocks that compose it (\textsc{amzn} and \textsc{nflx}) also have aleatory patterns; in contrast,  Eurostoxx has a long-term memory and, paradoxically, from an efficiency viewpoint it resembles more the  \textsc{btc} than the other index, which includes more disruptive companies.

\renewcommand{\arraystretch}{1.3} 
\begin{table}[H]
	\fontsize{7pt}{7pt}\selectfont
	\centering
\begin{tabular}{llll}
	\toprule  Name & $N$ &  Hurst & $p$-value \\
	\midrule	\midrule    BTC & 919 & 0.64169 & $8.56 \times 10^{-5}$ \\
	 ETH &  919 & 0.63337 & $2.28 \times 10^{-5}$\\
	 BNB &  919 & 0.65587 & $6.20 \times 10^{-5}$ \\
	XRP &  919 & 0.56646 & 0.01623 \\
	 ADA &  919 & 0.65739 & $7.57\times 10^{-4}$ \\
	MATIC & 919 & 0.68630 & $1.48 \times 10^{-5}$ \\
	SOL & 919 & 0.67028 & $5.53 \times 10^{-4}$\\
	TRX & 919 & 0.59418 & $2.41 \times 10^{-3}$ \\
	 DOT & 919 & 0.63734 & $3.81 \times 10^{-4}$ \\
Silver & 653 & 0.50182 & 0.97300 \\
	Gold  & 652 & 0.54179 & 0.23721 \\
	Crude & 633 &  0.59764 & $3.29 \times 10^{-4}$ \\
	Wheat & 608 &  0.58424 & $5.67 \times 10^{-3}$ \\
	Bonds & 633 &  0.54451 & $2.41 \times 10^{-3}$ \\
	Nasdaq & 633 & 0.53580 & 0.32121 \\
	EuroSt. & 650 & 0.58617 & $3.96 \times 10^{-4}$ \\
	 AMZN & 633 & 0.45961 & 0.28361 \\
	NFLX & 633  & 0.62543 & 0.03299 \\
	\bottomrule
\end{tabular}
\caption{Hurst exponent between 20/8/2020 and 24/2/2023 with daily data.} \label{Hurst-1}
\end{table}

\vspace*{-0.3cm}

Notice that the aleatory behaviour of   \textsc{xrp} already gives a  hint that the  consensus methods may influence the efficiency of \textsc{cc}. Indeed, \textsc{xrp} cannot be regarded as decentralised, since most of its nodes are controlled by a single company. This is because its aim is not to replace banks as intermediaries, but to improve their effectiveness,  as a substitute for the banking \textsc{swift}. 

Next, to discern if \textsc{trx} and \textsc{bnb}, which are neither fully decentralised, also display random patterns, we will perform a more refined fractal analysis, considering different time periods and/or modifying the time frequencies. Actually, when varying the latter, the outcomes in Table \ref{Hurst-2} reveal that \textsc{btc}'s movements are non-efficient at any scale, also exhibiting long-term memory. In contrast, \textsc{xrp} and \textsc{trx} operate mostly in an efficient way.

\renewcommand{\arraystretch}{1.2} 
\begin{table}[H]
	\fontsize{7pt}{7pt}\selectfont
	\centering
\begin{tabular}{lll}
	\toprule & Hurst & $p$-value \\
	\midrule	\midrule  {\bf BTC} & & \\
	15 min. & 0.53736 &  $1.12 \times 10^{-4}$ \\
	1 hour & 0.55226 & $9.92 \times 10^{-5}$ \\
	Daily & 0.64169 & $8.56 \times 10^{-5}$ \smallskip \\
	 {\bf  XRP} & & \\
	15 min. & 0.51401 & 0.06970 \\
	1 hour & 0.52551 & 0.01164 \\
	Daily & 0.56646 & 0.01623 \smallskip \\ 
	{\bf TRX} & & \\
	15 min. & 0.51740 & 0.04995 \\
	1 hour & 0.52192 & 0.04746 \\
	Daily & 0.59418 & $2.41 \times 10^{-3}$ \smallskip \\
	{\bf BNB} & & \\
	15 min. & 0.55127 & $1.11 \times 10^{-6}$ \\
	1 hour & 0.56468 & $1.30 \times 10^{-5}$ \\
	Daily & 0.65587 & $6.20 \times 10^{-5}$ \\
	\bottomrule
\end{tabular}
\caption{Hurst exponent: BTC, XRP, TRX and BNB for the full period but different time frequencies: 15 min. ($N = 88,\!060$ observations), 1 hour ($N = 22,\!019$) and daily ($N = 919$).} \label{Hurst-2}
\end{table}


But to observe traits of aleatoriness in \textsc{bnb}, we still need to reduce the time arc (or zoom in) to the 2nd subperiod (see Table \ref{Hurst-3}). As before,   although using daily splits \textsc{trx}  exhibits long-term memory, if we take high frequency data, it behaves following a random  motion, which is an early clue of lack of self-similarity. 

\renewcommand{\arraystretch}{1.2} 
\begin{table}[H]
	\fontsize{7pt}{7pt}\selectfont
	\centering
	\begin{tabular}{lll}
		\toprule & Hurst & $p$-value \\
		\hline	\hline & \\[-1.5ex]  {\bf XRP} & & \\
		15 min. & 0.51450 & 0.14754 \\
		1 hora & 0.51419 & 0.28596 \\
		Daily & 0.56816 & 0.13156 \smallskip \\
		{\bf TRX} & & \\
		15 min. & 0.51719 & 0.07521 \\
		1 hour & 0.52535 & 0.04041 \\
		Daily & 0.62667 & $1.35 \times 10^{-3}$ \smallskip \\
		{\bf BNB} & & \\
		15 min. & 0.53628 & $1.04 \times 10^{-3}$ \\
		1 hour & 0.53901 & $9.70 \times 10^{-3}$ \\
		Daily & 0.62968 & $0.04020$ \\
		\bottomrule
	\end{tabular}
	\caption{Hurst exponent: XRP, TRX and BNB for the 2nd subperiod (1/7/2022 -- 24/2/2023) and different time frequencies: 15 min. ($N = 22,\!869$ observations), 1 hour ($N = 5,\!718$) and daily ($N = 239$).} \label{Hurst-3}
\end{table}

\vspace*{-0.3cm}

Regarding the other six \textsc{cc}, whose algorithms are completely decentralised, they show persistent memory  in the full period (Table \ref{Hurst-A1-full}), and in both sub-intervals (Tables \ref{Hurst-A1-1st} and \ref{Hurst-A1}). Moreover, this remains invariant at any time scale, so we can claim that they have a consistent inefficient behaviour.



\subsection{Fractal features: does the graph change when zooming in?}

Next we inspect whether the return graphs of \textsc{btc} and other assets exhibit self similarity features, that is, if they look the same on different time scales. For this to hold, the efficiency character must not vary at different subintervals or if we change the frequency to take the data. 

Unlike \textsc{btc}, which moves a persistently at any frequency (cf.~Table \ref{Hurst-2}), silver and  Eurostoxx (Table \ref{Hurst-5})  change pattern as we reduce the partition. Specifically, for daily data silver moves randomly, but it is mean-reverting for high frequencies. Due to this shift, we deduce that it does not behave as a fractal. 
\renewcommand{\arraystretch}{1.2} 
\begin{table}[H]
	\fontsize{7pt}{7pt}\selectfont
	\centering
	\begin{tabular}{lcll}
		\toprule & N & Hurst & $p$-value \\
		\midrule	\midrule   {\bf Silver} & & \\
		15 min. & 60,320 & 0.42234 & $3.82 \times 10^{-5}$ \\
		1 hour & 15,463  & 0.42702 & $4.03 \times 10^{-4}$ \\
		Daily & 653  & 0.50182 & 0.97286 \smallskip \\
		{\bf Eurostoxx } & & \\
		15 min. & 23,376 & 0.45768 & $1.43 \times 10^{-3}$ \\
		1 hour & 5,849 & 0.48318 & 0.06186 \\
		Daily & 650 & 0.58617 & $3.96 \times 10^{-4}$ \\
		\bottomrule
	\end{tabular}
	\caption{Hurst exponent: silver and Eurostoxx between 20/8/2020 \& 24/2/2023 and different time frequencies.} \label{Hurst-5}
\end{table}

\vspace*{-0.3cm}

Moreover, one can also observe that traditional assets display a change in the memory character for different periods (see Table \ref{Hurst-A1-other}). Therefore, none of them can be regarded as a fractal object.

\renewcommand{\arraystretch}{1.2} 
\begin{table}[H]
	\fontsize{7pt}{7pt}\selectfont
	\centering
	\begin{tabular}{lllllll}
		\toprule
		& \multicolumn{2}{c}{20/8/2020 -- 24/2/2023} &  \multicolumn{2}{c}{20/8/2020 -- 1/7/2022}  &  \multicolumn{2}{c}{1/7/2022 -- 24/02/2023}  \\ \midrule 
		~ & Hurst & $p$-value & Hurst & $p$-value & Hurst & $p$-value \\ \midrule \midrule
		Silver & \textcolor{magenta}{0.5018} & 0.97286 & \textcolor{magenta}{0.5746} & 0.04957 & \textcolor{blue}{0.6340} & $2.81 \times 10^{-3}$  \\ 
		Gold & \textcolor{magenta}{0.5418} & 0.23721 & \textcolor{blue}{0.5880} & $3.82 \times 10^{-3}$  & \textcolor{magenta}{0.6688} & 0.02832 \\ 
		Crude & \textcolor{blue}{0.5976} & $3.29 \times 10^{-4}$  & \textcolor{magenta}{0.5561} & 0.03144 & \textcolor{magenta}{0.4498} & 0.38206 \\ 
		Wheat & \textcolor{blue}{0.5842} & $5.67 \times 10^{-3}$  & \textcolor{blue}{0.5861} & $5.36 \times 10^{-3}$  & \textcolor{magenta}{0.5647} & 0.23887 \\ 
		US bonds & \textcolor{blue}{0.5445} & $2.41 \times 10^{-3}$  & \textcolor{magenta}{0.5631} & 0.02122 & \textcolor{blue}{0.6406} & $3.43 \times 10^{-3}$  \\ 
		Nasdaq & \textcolor{magenta}{0.5358} & 0.32121 & \textcolor{magenta}{0.5835} & 0.03252 & \textcolor{magenta}{0.6475} & 0.01860 \\ 
		Eurostoxx & \textcolor{blue}{0.5862} & $3.96 \times 10^{-4}$  & \textcolor{blue}{0.6135} & $1.19 \times 10^{-3}$  & \textcolor{magenta}{0.6768} & 0.01810 \\ \bottomrule
	\end{tabular}
	\caption{Hurst exponent: traditional assets for different time periods and daily frequency. With colors we mark \textcolor{magenta}{random} and \textcolor{blue}{long memory} behaviour.} \label{Hurst-A1-other}
\end{table}

\begin{figure}[H] 
	\centering
	\includegraphics[scale=0.35]{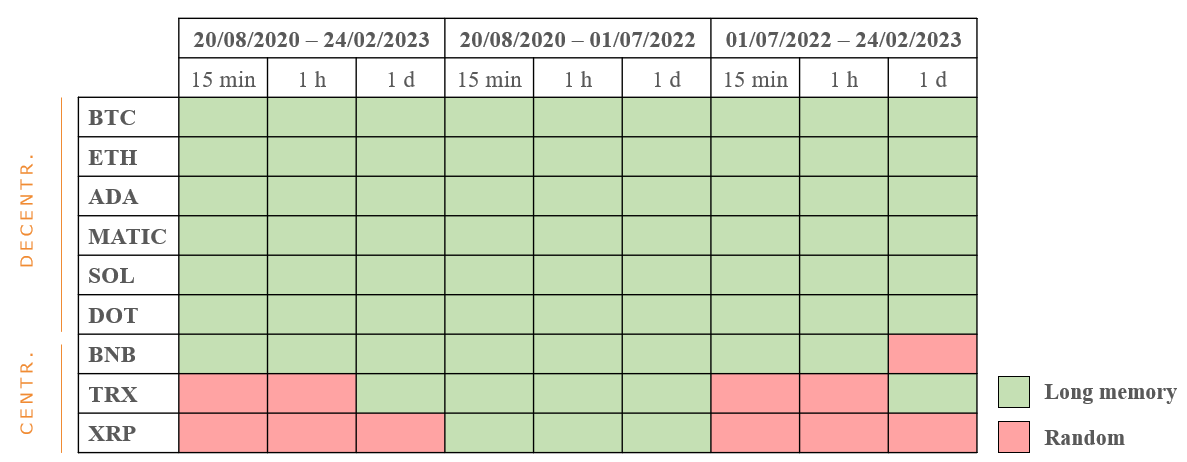}  \caption{Illustration of the self-similarity of \textsc{cc} for different time frames and frequencies (made with the outcomes from Tables \ref{Hurst-A1-full}, \ref{Hurst-A1-1st} and \ref{Hurst-A1}). Coloured at 0.1\% significance.} \label{color-cc}
\end{figure}

\vspace*{-0.3cm}
In contrast,  the scale-invariance  of \textsc{btc}  is shared by the five \textsc{cc}  whose consensus protocols are fully decentralised (see Figure \ref{color-cc}). This holds both in the complete study period and in the two sub-intervals considered, thus confirming the fractal nature of \textsc{cc} that operate peer-to-peer. Furthermore, this corroborates our guess that  the underlying consensus protocol is key to determine the efficiency (instead of the capitalization  as claimed in previous literature, since \textsc{bnb} and \textsc{xrp} are 3rd and 4th in volume, respectively; recall Table \ref{pareto-table}).

\begin{figure}[H] 
	\centering
	\includegraphics[scale=0.4]{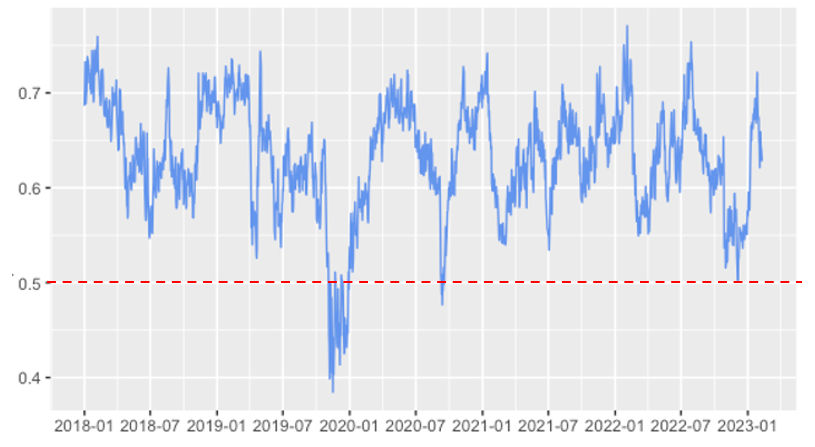}  \caption{Evolution of the Hurst exponent of BTC over time} \label{time-Hurst:fig}
\end{figure}

\vspace*{-0.3cm}
Additionally, we have analysed how the \textsc{btc}'s Hurst values vary as a function of time from March 2018 to Feb.~2023 with daily opening data. To address this question and calculate $H(t)$, a 150-value rolling window has been used. The outcomes (see Figure \ref{time-Hurst:fig}) show that \textsc{btc} presents long-term memory most of the time with $H$ values between 0.55 and 0.75, except at specific moments such as November 2019, when sharp oscillations within the anti-persistence zone occur.
\vspace*{-0.3cm}
\noindent  Notice that in the last trimester of 2022 the dynamics seems quite random, which could be influenced by the macroeconomomic situation tensioned by high inflation and rising interest rates.

\subsection{A striking turnabout: efficiency may happen (under a change of variable)}

We were puzzled by the claim in \cite{Nada} that if, instead of working directly with the  \textsc{btc} returns defined via \eqref{return}, we take the values $R_t^{17}$, then the resulting series becomes more efficient, except for the fact that their tests do not provide enough evidence to reject independence. As the authors restrict to \textsc{btc} and do not compute the Hurst exponent, we decided to update their study and broaden it including the variety of \textsc{cc} and traditional assets we are dealing with.

\begin{figure}[H] 
	\centering
	\includegraphics[scale=0.35]{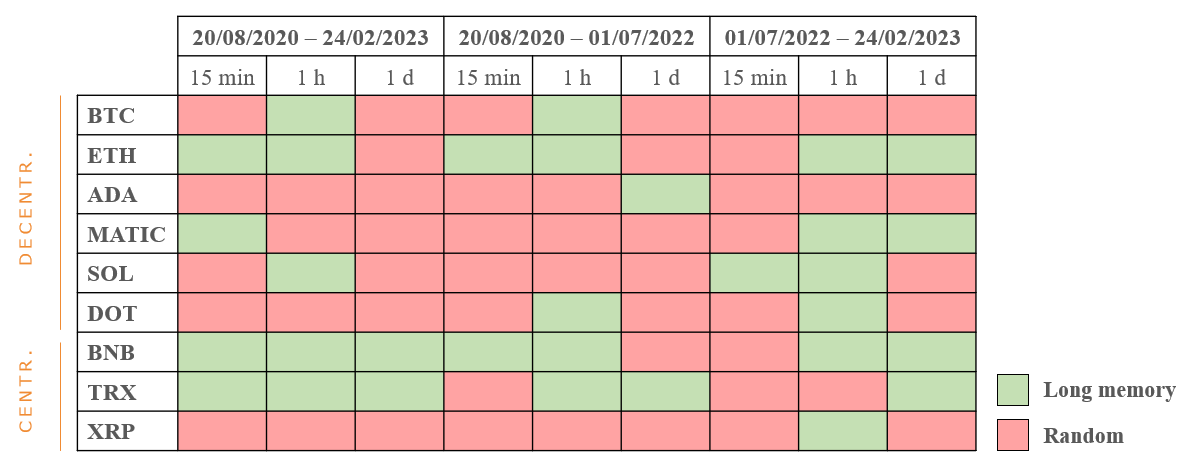}  \caption{Illustration of the self-similarity of \textsc{cc} for different time frames and frequencies (with the values of the return series raised to 17). Coloured at 0.1\% significance.} \label{color-cc-17}
\end{figure}

A quick glance at Figure \ref{color-cc-17} compared to Figure \ref{color-cc} already reveals that this simple transformation distorts the conclusions about efficiency, and unlike \cite{Nada} we also get that for most time frames and scales the new values behave independently. Furthermore, if we relax the significance requirements to 1\%, most of the table will be coloured as random. 

\vspace*{-0.3cm}
Nevertheless, our outcomes also reveal that the reply to the change of varia\-ble and shift to higher efficiency is not consistent, as \textsc{trx} moves oppositely towards long-memory character. Additionally, we lose information, since the new outcomes do not evidence any coherent common trend or difference according to neither technology nor liquidity.

Interestingly, traditional assets are more resilient to changes in the efficiency character in a coherent way (compare Table \ref{Hurst-A17-other} and Table \ref{Hurst-A1-other}), which gives a further support from an alternative perspective to our conclusion about the different nature of \textsc{cc} versus traditional assets.
\renewcommand{\arraystretch}{1.2} 
\begin{table}[H]
	\fontsize{7pt}{7pt}\selectfont
	\centering
	\begin{tabular}{lllllll}
		\toprule
		& \multicolumn{2}{c}{20/8/2020 -- 24/2/2023} &  \multicolumn{2}{c}{20/8/2020 -- 1/7/2022}  &  \multicolumn{2}{c}{1/7/2022 -- 24/02/2023}  \\ \midrule 
		~ & Hurst & $p$-value & Hurst & $p$-value & Hurst & $p$-value \\ \midrule \midrule
		Silver & \textcolor{magenta}{0.5213} & 0.0627 & \textcolor{magenta}{0.5255} & 0.0434 & \textcolor{magenta}{0.5654} & 0.0264  \\ 
		Gold & \textcolor{magenta}{0.5291} & 0.0971 & \textcolor{magenta}{0.5273} & 0.0205  & \textcolor{blue}{0.6534} & $2.68 \times 10^{-3}$ \\ 
		Crude & \textcolor{magenta}{0.4845} & 0.345  & \textcolor{magenta}{0.4847} & 0.4560 & \textcolor{blue}{0.6208} & $8.55 \times 10^{-5}$ \\ 
		Wheat & \textcolor{blue}{0.5241} & $3.76 \times 10^{-3}$  & \textcolor{blue}{0.5329} & $7.32 \times 10^{-3}$  & \textcolor{magenta}{0.5542} & 0.0205 \\ 
		US bonds & \textcolor{blue}{0.5600} & $7.94 \times 10^{-4}$  & \textcolor{magenta}{0.5913} & 0.0140 & \textcolor{magenta}{0.5586} & 0.0709  \\ 
		Nasdaq & \textcolor{magenta}{0.5203} & 0.2530 & \textcolor{blue}{0.5814} & $9.32 \times 10^{-5}$ & \textcolor{magenta}{0.5603} & 0.0102 \\ 
		Eurostoxx & \textcolor{blue}{0.5928} & $3.86 \times 10^{-4}$  & \textcolor{blue}{0.5977} & $4.49 \times 10^{-4}$  & \textcolor{magenta}{0.5571} & 0.0154 \\ \bottomrule
	\end{tabular}
	\caption{Hurst exponent: traditional assets for various  periods and daily frequency (with  return series raised to 17). Colors  mark \textcolor{magenta}{random} and \textcolor{blue}{long memory} behaviour.} \label{Hurst-A17-other}
\end{table}

\vspace*{-0.3cm}
For the purposes of the present paper, we leave this subsection as a matter of curiosity that deserves further and closer inspection, as we disagree with the premature statement in \cite{Nada} that any odd power will produce similar results. Indeed, very recent research \cite{Yu} proves that for a return series with $R_t > 0$ for all $t$, the fractional dimension keeps invariant under the change $R_t^q$ for every $q$.

It is actually an interesting future research line to find an optimal change of variable that turns any time series into a random Brownian motion, as this will enable to use the plethora of 
statistical techniques  based on the normal distribution, and drive  conclusions after undoing the transformation.
However, despite some numerical experiments can be helpful, a rigorous approach to this problem should wait for extra advances in fractal geometry, as it is an open problem to find a concrete formula of how the dimension would vary.

\subsection{Interdependence of Bitcoin and other assets}

To determine whether a diversified portfolio can be built with \textsc{cc} alone or they should be combined with other types of products, we obtain the Pearson correlations between the different assets  in the whole period and in both sub-intervals.  In view of the outcomes (see Table \ref{correl-comp}),  the second option happens, as \textsc{btc} has very strong positive correlations with all \textsc{cc} except \textsc{matic}.

\renewcommand{\arraystretch}{1.2} 
\begin{table}[H]
	\fontsize{7pt}{7pt}\selectfont
	\centering
\begin{tabular}{llll}
	\toprule & full period & 1st subperiod & 2nd subperiod \\ \midrule \midrule
	 BNB & $0.7318^{***}$ & $0.8217^{***}$ & $0.5129^{***}$ \\
	 TRX & $0.7089^{***}$ & $0.7785^{***}$ & $0.9059^{***}$ \\
	 SOL & $0.6606^{***}$ & $0.6259^{***}$ & $0.6932^{***}$ \\
	 MATIC & $0.5037^{***}$ & $0.5779^{***}$ & $0.4791^{***}$ \\
	 ETH & $0.8422^{***}$ & $0.8342^{***}$ & $0.8396^{***}$ \\
	 DOT & $0.9527^{***}$ & $0.9380^{***}$ & $0.8011^{***}$ \\
	 ADA & $0.8275^{***}$ & $0.7851^{***}$ & $0.7125^{***}$ \\
	 XRP & $0.8050^{***}$ & $0.7656^{***}$ & 0.0441 \\
	 TSLA & $0.6240^{***}$ & $0.6812^{***}$ & $0.4635^{***}$ \\
	 NFLX & $0.5564^{***}$ & $0.3544^{***}$ & -0.006 \\
	 AMZN & $0.5540^{***}$ & $0.2980^{***}$ & $0.5396^{***}$ \\
	 KO & -$0.1100^{**}$ & $0.2099^{***}$ & 0.0053 \\
	 PG & $0.1017^*$ & $0.0924^*$ & -$0.1615^*$ \\
	 JNJ & $0.2074^{***}$ & $0.4573^{***}$ & -$0.5148^{***}$ \\
	 Crude & $0.1331^{***}$ & $0.3493^{***}$ & $0.3249^{***}$ \\
	 Wheat & $0.1005^*$ & $0.2084^{***}$ & -0.0784 \\
	 US bonds & $0.2886^{***}$ & -$0.2597^{***} $ & $0.4490^{***}$ \\
	 Silver & $0.3457^{***}$ & 0.012 & -$0.1935^*$ \\
	 Gold & -$0.1281^{**}$ & -$0.5741^{***}$ & $0.2076^{**}$ \\
	 Nasdaq & $0.8458^{***}$ & $0.7984^{***}$ & $0.6902^{***}$ \\
	 EuroStoxx & $0.6749^{***}$ & $0.8005^{***}$ & 0.1553 \\
	\bottomrule
\end{tabular}
\caption{Correlation of daily returns (for weekdays) between \textsc{btc} and other assets: period from 20-08-2020 to 24/2/2023, and the sub-intervals from 20/8/2020 to 1/7/2022 and from 1/7/2022 to 24/2/2023. $^{***}$ ($^{**}$, $^*$) denotes significance at 0.1\% (1\%, 5\%) significance level.} \label{correl-comp}
\end{table}

\vspace*{-0.3cm}

We also confirm our previous guess that growth stocks (\textsc{tsla, nflx} \& \textsc{amzn}, and hence the Nasdaq index) have similar dynamics than \textsc{btc}. However, it has a weak linear relation with value stocks, hence the mix of these with \textsc{btc} has no diversification power, and it is a clever strategy to combine them with e.g. \textsc{nflx}, with whom strong negative correlations pop up (see Table \ref{correl-2-full}). Moreover, we conclude that \textsc{btc}  is far from acting like the more conservative safe haven products for investors, as no no correlation with them arose.

If we now restrict to the interrelations within the {\it cryptosphere}, all \textsc{cc} are very tightly correlated as expected (see Tables \ref{correl-CC-f} and \ref{correl-CC-1} for the full matrices of the full period and the first subinterval, respectively). Nevertheless, it is quite surprising that, if we focus on the last subperiod, the completely unprecedented event of negative correlations between them emerges (cf.~Table \ref{correl-CC-2}).

If we move from the whole time frame to the earlier subset, most of the values match up to the 1st decimal, except some shifts in the interdependences of gold and silver with growth stocks (see Tables \ref{correl-2-full} and \ref{correl-2-1st}). But if we look at  the last subinterval, unexpectedly the landscape changes quite radically. Certainly, \textsc{btc} loses correlation strength  with growth stocks (cf.~Table \ref{correl-2-2nd}); in parallel, slight negative correlations start to emerge with assets such as \textsc{jnj} or silver.

\section{Conclusions} \label{conclu}

Our main goal was to decide whether the evolution of \textsc{btc} prices is random (as assumed by the \textsc{emh}) or follows chaotic (but more predictable) patterns, opening the door to forecasting opportunities. Via the Hurst values ($H$), we confirm that the \textsc{btc} dynamics are not independent, but have long memory,  even if we change the time interval or the frequency of data collection.	The latter indicates that the graphs of \textsc{btc} returns are self similar, an essential feature to confirm their nature as fractals. 

On the contrary, when changing the scale, we get efficiency shifts for traditional assets, as well as those \textsc{cc}  whose  consensus protocols have centralised features (\textsc{bnb, trx} {\small \&} \textsc{xrp}). Up to our knowledge, we pioneer in stressing that it is the underlying technology (intead of the liquidity) the key to determine differences between the performance of \textsc{cc} within the market.


As a corollary, the fractality of \textsc{btc}, characterised by chaotic events occurring in waves rather than isolatedly, contradicts every conservative investor's desire for it to be a store of value. This invalidates that \textsc{btc} could be considered a safe haven, while not holding the narrative that it is the {\it digital gold}.

From our correlational study,   \textsc{btc} generally exhibits the same trend as other \textsc{cc} and growth stocks. In practice, this implies that it is not possible to build a risk-controlled portfolio with this type of assets alone.
At the same time, there is no tight correlation of \textsc{btc} with {\it conservative} products; consequently, it is also very difficult exploit \textsc{btc} as a diversification tool.

Complementarily, if we focus on the later subperiod, the correlations of all assets get disrupted. It will be interesting for future work to analyse whether this paradigm shift is an isolated event or is sustained over time.
A plausible justification within the actual context could come from the influence of a sharp rise in interest rates, joint with a global economy strained by inflation.  

To recap, we  detect a high degree of persistence in \textsc{btc} prices. This may be due to the lack of investor confidence in this market, as it is unstructured and lacks oversight by the authorities. With more legislative certainty, participation will increase, which may lead to a rise of efficiency (or decrease in predictability) that would stabilise the {\it cryptoeconomy}.

\vspace*{-0.4cm}

\appendix

\section{Table giving a literature overview}

\vspace*{-2.0cm}
\begin{figure}[H]
\centering
\hspace*{-1cm} \includegraphics[scale=0.55]{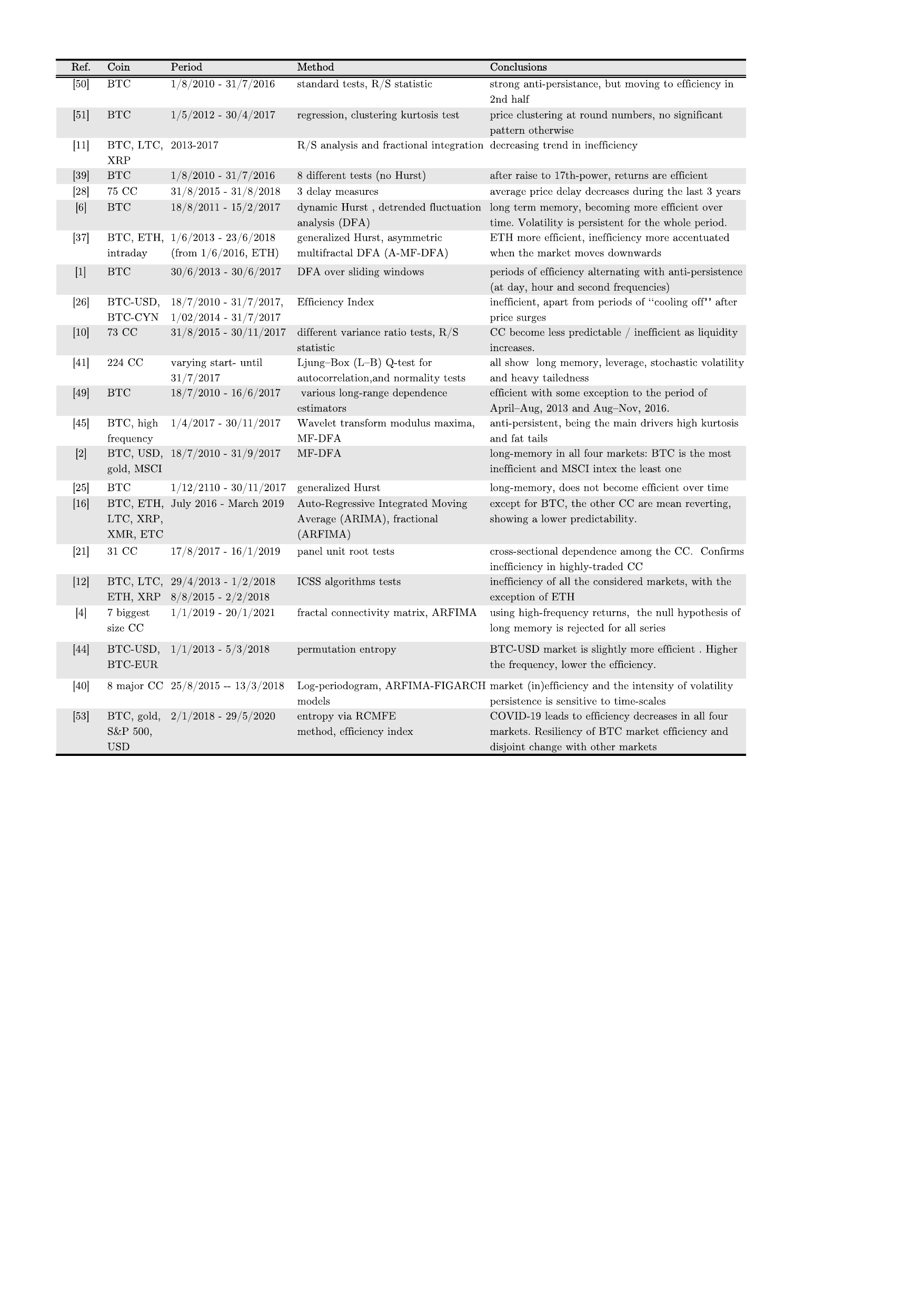}	\vspace*{-10.3cm}\caption{Overview of literature about efficiency/persistence} \label{lit:table}
\end{figure}

\vspace*{-0.4cm}
\section{Complete tables with Hurst exponents}

\vspace*{-0.4cm}
\renewcommand{\arraystretch}{1.2} 

\begin{table}[H] 
	\fontsize{5pt}{5pt}\selectfont
	\centering
	\begin{tabular}{
			@{}
			p{\dimexpr1.0\textwidth}
		}
		\toprule
		\begin{tabular}{lll}
			& Hurst & $p$-value  \vspace*{-0.025cm}\\
			\midrule   \vspace*{-0.04cm} {\bf  BTC} & & \\
			15 min. &0.5374 & $1.12 \times 10^{\scaleto{-4}{3.5pt}}$ 
			\\
			1 hour & 0.5523 & $9.92 \times 10^{\scaleto{-5}{3.5pt}}$  \\
			Daily & 0.6417 & $8.56 \times 10^{\scaleto{-5}{3.5pt}}$ \\
			& & \\
			{\bf ETH} & & \\
			15 min. & 0.5375 & $2.87 \times 10^{\scaleto{-6}{3.5pt}}$ \\
			1 hour & 0.5525 & $5.99 \times 10^{\scaleto{-7}{3.5pt}}$ \\
			Daily & 0.6334 & $2.28 \times 10^{\scaleto{-5}{3.5pt}}$ \\
			& & \\
			{\bf BNB} & & \\
			15 min. & 0.5513 & $1.11 \times 10^{\scaleto{-6}{3.5pt}}$ \\
			1 hour & 0.5647 & $1.30 \times 10^{\scaleto{-5}{3.5pt}}$ \\
			Daily & 0.6559 & $6.20 \times 10^{\scaleto{-5}{3.5pt}}$ 
		\end{tabular}\hfill
		\begin{tabular}{lll}
			& Hurst & $p$-value \\
			\midrule   {\bf  XRP} & & \\
			15 min. & 0.5140 & 0.06970 
			\\
			1 hour & 0.5219 & 0.04746  \\
			Daily & 0.5942 & $2.41 \times 10^{\scaleto{-3}{3.5pt}}$  \\
			& & \\
			{\bf ADA} & & \\
			15 min. & 0.5356 & $1.52 \times 10^{\scaleto{-3}{3.5pt}}$ \\
			1 hour & 0.5540 & $8.30 \times 10^{\scaleto{-4}{3.5pt}}$ \\
			Daily & 0.6574 & $7.57 \times 10^{\scaleto{-4}{3.5pt}}$ \\
			& & \\
			{\bf MATIC} & & \\
			15 min. & 0.5539 & $3.77 \times 10^{\scaleto{-5}{3.5pt}}$ \\
			1 hour & 0.5767 & $3.56 \times 10^{\scaleto{-5}{3.5pt}}$ \\
			Daily & 0.6863 & $1.48 \times 10^{\scaleto{-5}{3.5pt}}$ 
		\end{tabular}\hfill
		\begin{tabular}{lll}
			& Hurst & $p$-value \\
			\midrule   {\bf  SOL} & & \\
			15 min. & 0.5480 & $1.45 \times 10^{\scaleto{-4}{3.5pt}}$ 
			\\
			1 hour & 0.5632 & $2.39 \times 10^{\scaleto{-4}{3.5pt}}$  \\
			Daily & 0.6703 & $5.53 \times 10^{\scaleto{-4}{3.5pt}}$  \\
			& & \\
			{\bf TRX} & & \\
			15 min. &0.5174 & 0.04995 \\
			1 hour & 0.5219 & 0.04746 \\
			Daily &0.5942 & $2.41 \times 10^{\scaleto{-3}{3.5pt}}$ \\
			& & \\
			{\bf DOT} & & \\
			15 min. & 0.5276 & $8.51 \times 10^{\scaleto{-4}{3.5pt}}$ \\
			1 hour & 0.5413 & $5.00 \times 10^{\scaleto{-4}{3.5pt}}$ \\
			Daily & 0.6373 & $3.81 \times 10^{\scaleto{-4}{3.5pt}}$ 
		\end{tabular}
		\\
		\bottomrule
	\end{tabular}
\vspace*{-0.2cm}	\caption{Hurst exponent: cryptocurrencies between 20/8/2020  and 1/7/2022.} \label{Hurst-A1-full}
\end{table}

\vspace*{-0.4cm}

\renewcommand{\arraystretch}{1.2} 

\begin{table}[H] 
	\fontsize{5pt}{5pt}\selectfont
	\centering
	\begin{tabular}{
			@{}
			p{\dimexpr1.0\textwidth}
		}
		\toprule
	\begin{tabular}{lll}
			 & Hurst & $p$-value  \vspace*{-0.025cm}\\
		\midrule  \vspace*{-0.04cm} {\bf  BTC} & & \\
		15 min. &0.5462 & $1.62 \times 10^{\scaleto{-6}{3.5pt}}$ 
		\\
		1 hour & 0.5651 & $6.70 \times 10^{\scaleto{-7}{3.5pt}}$  \\
		Daily & 0.6670 & $1.20 \times 10^{\scaleto{-5}{3.5pt}}$ \\
		& & \\
		{\bf ETH} & & \\
		15 min. & 0.5450 & $1.72 \times 10^{\scaleto{-5}{3.5pt}}$ \\
		1 hour & 0.5637 & $9.66 \times 10^{\scaleto{-6}{3.5pt}}$ \\
		Daily & 0.6685 & $1.43 \times 10^{\scaleto{-4}{3.5pt}}$ \\
		& & \\
		{\bf BNB} & & \\
		15 min. & 0.5569 & $1.20 \times 10^{\scaleto{-6}{3.5pt}}$ \\
		1 hour & 0.5736 & $5.70 \times 10^{\scaleto{-6}{3.5pt}}$ \\
		Daily & 0.6622 & $4.05 \times 10^{\scaleto{-4}{3.5pt}}$ 
	\end{tabular}\hfill
	\begin{tabular}{lll}
		 & Hurst & $p$-value \\
		\midrule   {\bf  XRP} & & \\
		15 min. & 0.5271 & $5.88 \times 10^{\scaleto{-4}{3.5pt}}$ 
		\\
		1 hour & 0.5446 & $3.75 \times 10^{\scaleto{-5}{3.5pt}}$  \\
		Daily & 0.5820 & $3.52 \times 10^{\scaleto{-3}{3.5pt}}$  \\
		& & \\
		{\bf ADA} & & \\
		15 min. & 0.5411 & $2.04 \times 10^{\scaleto{-4}{3.5pt}}$ \\
		1 hour & 0.5630 & $8.95 \times 10^{\scaleto{-5}{3.5pt}}$ \\
		Daily & 0.6733 & $8.42 \times 10^{\scaleto{-4}{3.5pt}}$ \\
		& & \\
		{\bf MATIC} & & \\
		15 min. & 0.5505 & $6.43 \times 10^{\scaleto{-4}{3.5pt}}$ \\
		1 hour & 0.5752 & $4.11 \times 10^{\scaleto{-4}{3.5pt}}$ \\
		Daily & 0.6918 & $2.85 \times 10^{\scaleto{-3}{3.5pt}}$ 
	\end{tabular}\hfill
	\begin{tabular}{lll}
		 & Hurst & $p$-value \\
		\midrule   {\bf  SOL} & & \\
		15 min. & 0.5627 & $5.13 \times 10^{\scaleto{-6}{3.5pt}}$ 
		\\
		1 hour & 0.5819 & $9.38 \times 10^{\scaleto{-6}{3.5pt}}$  \\
		Daily & 0.7275 & $5.69 \times 10^{\scaleto{-5}{3.5pt}}$  \\
		& & \\
		{\bf TRX} & & \\
		15 min. &0.5310 & $6.40 \times 10^{\scaleto{-4}{3.5pt}}$ \\
		1 hour & 0.5384 & $8.33 \times 10^{\scaleto{-4}{3.5pt}}$ \\
		Daily &0.5949 & $1.42 \times 10^{\scaleto{-3}{3.5pt}}$ \\
		& & \\
		{\bf DOT} & & \\
		15 min. & 0.5352 & $6.60 \times 10^{\scaleto{-6}{3.5pt}}$ \\
		1 hour & 0.5528 & $2.58 \times 10^{\scaleto{-6}{3.5pt}}$ \\
		Daily & 0.6398 & $3.11 \times 10^{\scaleto{-4}{3.5pt}}$ 
	\end{tabular}
\\
\bottomrule
\end{tabular}
\vspace*{-0.3cm}	\caption{Hurst exponent: cryptocurrencies between 20/8/2020  and 1/7/2022.} \label{Hurst-A1-1st}
\end{table}

\vspace*{-0.4cm}

\renewcommand{\arraystretch}{1.2} 

\begin{table}[H] 
	\fontsize{5pt}{5pt}\selectfont
	\centering
	\begin{tabular}{
			@{}
			p{\dimexpr1.0\textwidth}
		}
		\toprule
		\begin{tabular}{lll}
			& Hurst & $p$-value  \vspace*{-0.02cm}\\
			\midrule  \vspace*{-0.04cm} {\bf  BTC} & & \\
			15 min. &0.5406 & $1.36 \times 10^{\scaleto{-4}{3.5pt}}$ 
			\\
			1 hour & 0.5508 & $2.39 \times 10^{\scaleto{-4}{3.5pt}}$  \\
			Daily & 0.6227 & $4.60 \times 10^{\scaleto{-3}{3.5pt}}$ \\
			& & \\
			{\bf ETH} & & \\
			15 min. & 0.5491 & $1.41 \times 10^{\scaleto{-7}{3.5pt}}$ \\
			1 hour & 0.5641 & $6.68 \times 10^{\scaleto{-7}{3.5pt}}$ \\
			Daily & 0.6645 & $9.69 \times 10^{\scaleto{-4}{3.5pt}}$ \\
			& & \\
			{\bf BNB} & & \\
			15 min. & 0.5363 & $1.04 \times 10^{\scaleto{-3}{3.5pt}}$ \\
			1 hour & 0.5390 & $9.70 \times 10^{\scaleto{-3}{3.5pt}}$ \\
			Daily & 0.6297 & $0.04020$ 
		\end{tabular}\hfill
		\begin{tabular}{lll}
			& Hurst & $p$-value \\
			\midrule   {\bf  XRP} & & \\
			15 min. & 0.5145 & 0.14754
			\\
			1 hour & 0.5141 & 0.28596\textcolor{white}{$ 10^{\scaleto{-5}{3.5pt}}$} \\
			Daily & 0.5682 & 0.13156 \\
			& & \\
			{\bf ADA} & & \\
			15 min. & 0.5376 & $1.46 \times 10^{\scaleto{-5}{3.5pt}}$ \\
			1 hour & 0.5531 & $1.56 \times 10^{\scaleto{-5}{3.5pt}}$ \\
			Daily & 0.6832 & $3.98 \times 10^{\scaleto{-5}{3.5pt}}$ \\
			& & \\
			{\bf MATIC} & & \\
			15 min. & 0.5377 & $4.22 \times 10^{\scaleto{-4}{3.5pt}}$ \\
			1 hour & 0.5506 & $5.50 \times 10^{\scaleto{-4}{3.5pt}}$ \\
			Daily & 0.6468 & $2.50 \times 10^{\scaleto{-3}{3.5pt}}$ 
		\end{tabular}\hfill
		\begin{tabular}{lll}
			& Hurst & $p$-value \\
			\midrule   {\bf  SOL} & & \\
			15 min. & 0.5604 & $1.03 \times 10^{\scaleto{-7}{3.5pt}}$ 
			\\
			1 hour & 0.5854 & $2.83 \times 10^{\scaleto{-7}{3.5pt}}$  \\
			Daily & 0.6767 & $1.85 \times 10^{\scaleto{-4}{3.5pt}}$  \\
			& & \\
			{\bf TRX} & & \\
			15 min. &0.5172 & 0.07521 \\
			1 hour & 0.5254 & 0.04041 \\
			Daily &0.6267 & $1.35 \times 10^{\scaleto{-3}{3.5pt}}$ \\
			& & \\
			{\bf DOT} & & \\
			15 min. & 0.5322 & $1.54 \times 10^{\scaleto{-3}{3.5pt}}$ \\
			1 hour & 0.5463 & $2.41 \times 10^{\scaleto{-3}{3.5pt}}$ \\
			Daily & 0.6829 & $2.74 \times 10^{\scaleto{-3}{3.5pt}}$ 
		\end{tabular}
		\\
		\bottomrule
	\end{tabular}
\vspace*{-0.3cm}	\caption{Hurst exponent: cryptocurrencies between 1/7/2022 and 24/2/2023.} \label{Hurst-A1}
\end{table}

\vspace*{-0.4cm}

\section{Correlation matrices for criptoassets and BTC versus traditional assets}

Hereafter  $^{***}$ ($^{**}$, $^*$) denotes significance at 0.1\% (1\%, 5\%) significance level.
\vspace*{-0.3cm}
\renewcommand{\arraystretch}{1.2} 
\begin{table}[H]
	\fontsize{7pt}{7pt}\selectfont
	\centering
	\resizebox{\textwidth}{!}{\begin{tabular}{llllllllll}
			\hline & \\[-1.5ex]
			&	BTC & BNB & TRX & SOL & MATIC & ETH & DOT & ADA & XRP \\
			\hline	\hline & \\[-1.5ex] 
			BTC & 1 & ~ & ~ & ~ & ~ & ~ & ~ & ~ & ~ \\ 
			BNB & 0.7318$^{***}$ & 1 & ~ & ~ & ~ & ~ & ~ & ~ & ~ \\ 
			TRX & 0.7089$^{***}$ & 0.8866$^{***}$ & 1 & ~ & ~ & ~ & ~ & ~ & ~ \\ 
			SOL & 0.6606$^{***}$ & 0.7489$^{***}$ & 0.5968$^{***}$ & 1 & ~ & ~ & ~ & ~ & ~ \\ 
			MATIC & 0.5037$^{***}$ & 0.8222$^{***}$ & 0.5711$^{***}$ & 0.7780$^{***}$ & 1 & ~ & ~ & ~ & ~ \\ 
			ETH & 0.8422$^{***}$ & 0.9057$^{***}$ & 0.7742$^{***}$ & 0.8825$^{***}$ & 0.8378*** & 1 & ~ & ~ & ~ \\ 
			DOT & 0.9527$^{***}$ & 0.7042$^{***}$ & 0.7252$^{***}$ & 0.6598$^{***}$ & 0.4609*** & 0.8008$^{***}$ & 1 & ~ & ~ \\ 
			ADA & 0.8275$^{***}$ & 0.7456$^{***}$ & 0.7379$^{***}$ & 0.7017$^{***}$ & 0.6312$^{***}$ & 0.8524$^{***}$ & 0.8423$^{***}$ & 1 & ~ \\ 
			XRP & 0.8050$^{***}$ & 0.7997$^{***}$ & 0.8387$^{***}$ & 0.6221$^{***}$ & 0.5788$^{***}$ & 0.8152$^{***}$ & 0.8233$^{***}$ & 0.8548$^{***}$ & 1 \\ \hline
	\end{tabular}}
\vspace*{-0.3cm}	\caption{Correlation between cryptoassets between 20/8/2020 \& 24/2/2023} \label{correl-CC-f}
\end{table}

\vspace*{-0.4cm}

\renewcommand{\arraystretch}{1.2} 
\begin{table}[H]
	\fontsize{7pt}{7pt}\selectfont
	\centering
	\resizebox{\textwidth}{!}{\begin{tabular}{llllllllll}
			\hline & \\[-1.5ex]
			&	BTC & BNB & TRX & SOL & MATIC & ETH & DOT & ADA & XRP \\
			\hline	\hline & \\[-1.5ex] 
			BTC & 1 & ~ & ~ & ~ & ~ & ~ & ~ & ~ & ~ \\ 
			BNB & 0.8217$^{***}$ & 1 & ~ & ~ & ~ & ~ & ~ & ~ & ~ \\ 
			TRX & 0.7785$^{***}$ & 0.8917$^{***}$ & 1 & ~ & ~ & ~ & ~ & ~ & ~ \\ 
			SOL & 0.6259$^{***}$ & 0.7686$^{***}$ & 0.5974$^{***}$ & 1 & ~ & ~ & ~ & ~ & ~ \\ 
			MATIC & 0.5779$^{***}$ & 0.8250$^{***}$ & 0.5768$^{***}$ & 0.8175$^{***}$ & 1 & ~ & ~ & ~ & ~ \\ 
			ETH & 0.8342$^{***}$ & 0.9392$^{***}$ & 0.7925$^{***}$ & 0.8741$^{***}$ & 0.8827$^{***}$ & 1 & ~ & ~ & ~ \\ 
			DOT & 0.9380$^{***}$ & 0.7820$^{***}$ & 0.7933$^{***}$ & 0.6235$^{***}$ & 0.5258$^{***}$ & 0.7821$^{***}$ & 1 & ~ & ~ \\ 
			ADA & 0.7851$^{***}$ & 0.7998$^{***}$ & 0.7793$^{***}$ & 0.6684$^{***}$ & 0.6965$^{***}$ & 0.8371$^{***}$ & 0.8029$^{***}$ & 1 & ~ \\ 
			XRP & 0.7656$^{***}$ & 0.8538$^{***}$ & 0.8927$^{***}$ & 0.5820$^{***}$ & 0.6313$^{***}$ & 0.7979$^{***}$ & 0.7868$^{***}$ & 0.8290$^{***}$ & 1 \\ \hline
	\end{tabular}}
\vspace*{-0.3cm}	\caption{Correlation between \textsc{cc} for the 1st subperiod between 20/8/2020 \& 1/7/2022} \label{correl-CC-1}
\end{table}

\vspace*{-0.4cm}
\renewcommand{\arraystretch}{1.2} 
\begin{table}[H]
	\fontsize{7pt}{7pt}\selectfont
	\centering
	\resizebox{\textwidth}{!}{\begin{tabular}{llllllllll}
			\hline & \\[-1.5ex]
			&	BTC & BNB & TRX & SOL & MATIC & ETH & DOT & ADA & XRP \\
			\hline	\hline & \\[-1.5ex] 
			BTC & 1 & ~ & ~ & ~ & ~ & ~ & ~ & ~ & ~ \\ 
			BNB & 0.5929$^{***}$ & 1 & ~ & ~ & ~ & ~ & ~ & ~ & ~ \\ 
			TRX & 0.9059$^{***}$ & 0.2921$^{***}$ & 1 & ~ & ~ & ~ & ~ & ~ & ~ \\ 
			SOL & 0.6932$^{***}$ & 0.1368 & 0.8407$^{***}$ & 1 & ~ & ~ & ~ & ~ & ~ \\ 
			MATIC & 0.4791$^{***}$ & 0.7286$^{***}$ & 0.2211$^{**}$ & -0.1586$^{*}$ & 1 & ~ & ~ & ~ & ~ \\ 
			ETH & 0.8396$^{***}$ & 0.7048$^{***}$ & 0.6931$^{***}$ & 0.5532$^{***}$ & 0.5588$^{***}$ & 1 & ~ & ~ & ~ \\ 
			DOT & 0.8011$^{***}$ & 0.3992$^{***}$ & 0.8477$^{***}$ & 0.8809$^{***}$ & 0.1309 & 0.7540$^{***}$ & 1 & ~ & ~ \\ 
			ADA & 0.7125$^{***}$ & 0.2197$^{***}$ & 0.8101$^{***}$ & 0.9460$^{***}$ & -0.0652 & 0.6323$^{***}$ & 0.9413$^{***}$ & 1 & ~ \\ 
			XRP & 0.044$^{***}$ & 0.3717$^{***}$ & -0.002$^{***}$ & 0.090$^{***}$ & 0.1914$^{*}$ & 0.0062 & -0.0515 & -0.0496 & 1 \\ \hline
	\end{tabular}}
\vspace*{-0.3cm}	\caption{Correlation between \textsc{cc} for the 2nd subperiod between 1/7/2022 \& 24/2/2023} \label{correl-CC-2}
\end{table}

\vspace*{-0.4cm}
\renewcommand{\arraystretch}{1.3} 
\begin{table}[H]
	\fontsize{7pt}{7pt}\selectfont
	\centering
	\resizebox{\textwidth}{!}{
		\begin{tabular}{lllllllllllllll}
			\hline & \\[-1.5ex]
			&
			BTC & TSLA & NFLX & AMZN & KO & PG & JNJ & OIL & WHEAT & ZNM23 & SILVER & GOLD & CCMP & SX5E  \\ 	\hline	\hline & \\[-1.5ex]
			BTC &	1 & ~ & ~ & ~ & ~ & ~ & ~ & ~ & ~ & ~ & ~ & ~ & ~ & ~ \\ 
			TSLA & 	0.6240$^{***}$ & 1 & ~ & ~ & ~ & ~ & ~ & ~ & ~ & ~ & ~ & ~ & ~ & ~ \\ 
			NFLX &	0.5564$^{***}$ & 0.1255$^{**}$ & 1 & ~ & ~ & ~ & ~ & ~ & ~ & ~ & ~ & ~ & ~ & ~ \\ 
			AMZN &	0.5540$^{***}$ & 0.3264$^{***}$ & 0.8124$^{***}$ & 1 & ~ & ~ & ~ & ~ & ~ & ~ & ~ & ~ & ~ & ~ \\ 
			KO &	-0.1100$^{**}$ & 0.2883$^{***}$ & -0.7136$^{***}$ & -0.5929$^{***}$ & 1 & ~ & ~ & ~ & ~ & ~ & ~ & ~ & ~ & ~ \\ 
			PG &	0.1017$^{*}$ & 0.4000$^{***}$ & -0.1337$^{***}$ & -0.0511 & 0.6250$^{***}$ & 1 & ~ & ~ & ~ & ~ & ~ & ~ & ~ & ~ \\ 
			JNJ &	0.2074$^{***}$ & 0.3453$^{***}$ & -0.4702$^{***}$ & -0.3805$^{***}$ & 0.7655$^{***}$ & 0.4230$^{***}$ & 1 & ~ & ~ & ~ & ~ & ~ & ~ & ~ \\ 
			OIL &	0.1331$^{***}$ & 0.4739$^{***}$ & -0.6360$^{***}$ & -0.4502$^{***}$ & 0.8346$^{***}$ & 0.4178$^{***}$ & 0.7680$^{***}$ & 1 & ~ & ~ & ~ & ~ & ~ & ~ \\ 
			WHEAT &	0.1005$^{*}$ & 0.4453$^{***}$ & -0.5995$^{***}$ & -0.3983$^{***}$ & 0.7225$^{***}$ & 0.4370$^{***}$ & 0.6566$^{***}$ & 0.8853$^{***}$ & 1 & ~ & ~ & ~ & ~ & ~ \\ 
			ZNM23 &	0.2886$^{***}$ & -0.0723 & 0.8077$^{***}$ & 0.8527$^{***}$ & -0.7906$^{***}$ & -0.1962$^{***}$ & -0.6048$^{***}$ & -0.7307$^{***}$ & -0.6281$^{***}$ & 1 & ~ & ~ & ~ & ~ \\ 
			SILVER & 	0.3457$^{***}$ & -0.1752$^{***}$ & 0.6281$^{***}$ & 0.5899$^{***}$ & -0.5643$^{***}$ & -0.1753$^{***}$ & -0.2711$^{***}$ & -0.5135$^{***}$ & -0.4070$^{***}$ & 0.7052$^{***}$ & 1 & ~ & ~ & ~ \\ 
			GOLD &	-0.1281$^{**}$ & -0.1695$^{***}$ & 0.1380$^{***}$ & 0.1976$^{***}$ & -0.0755 & 0.3312$^{***}$ & -0.1653$^{***}$ & -0.1631$^{***}$ & 0.018 & 0.3385$^{***}$ & 0.5770$^{***}$ & 1 & ~ & ~ \\ 
			CCMP &	0.8458$^{***}$ & 0.6579$^{***}$ & 0.7100$^{***}$ & 0.7734$^{***}$ & -0.1598$^{***}$ & 0.2298$^{***}$ & 0.1021$^{*}$ & -0.0114 & -0.068 & 0.4794$^{***}$ & 0.3704$^{***}$ & 0.0016 & 1 & ~ \\ 
			SX5E&	0.6749$^{***}$ & 0.4388$^{***}$ & 0.3532$^{***}$ & 0.1908$^{***}$ & 0.2798$^{***}$ & 0.3701$^{***}$ & 0.4401$^{***}$ & 0.2734$^{***}$ & 0.1016$^{*}$ & -0.0676 & 0.1415$^{***}$ & -0.075 & 0.7065$^{***}$ & 1 \\ \hline
	\end{tabular}}
\vspace*{-0.3cm}	\caption{Correlation between BTC and traditional assets (20/8/2020 -- 24/2/2023)} \label{correl-2-full}
\end{table}

\vspace*{-0.4cm}

\renewcommand{\arraystretch}{1.3} 
\begin{table}[H]
	\fontsize{7pt}{7pt}\selectfont
	\centering
	\resizebox{\textwidth}{!}{
		\begin{tabular}{lllllllllllllll}
			\hline & \\[-1.5ex]
			&
			BTC & TSLA & NFLX & AMZN & KO & PG & JNJ & OIL & WHEAT & ZNM23 & SILVER & GOLD & CCMP & SX5E    \\ 	\hline	\hline & \\[-1.5ex]
			BTC &	1 & ~ & ~ & ~ & ~ & ~ & ~ & ~ & ~ & ~ & ~ & ~ & ~ & ~ \\ 
			TSLA & 	0.6812$^{***}$ & 1 & ~ & ~ & ~ & ~ & ~ & ~ & ~ & ~ & ~ & ~ & ~ & ~ \\ 
			NFLX &	0.3544$^{***}$ & 0.0853 & 1 & ~ & ~ & ~ & ~ & ~ & ~ & ~ & ~ & ~ & ~ & ~ \\ 
			AMZN &	0.2980$^{***}$ & 0.0984$^{*}$ & 0.8644$^{***}$ & 1 & ~ & ~ & ~ & ~ & ~ & ~ & ~ & ~ & ~ & ~ \\ 
			KO &	0.2099$^{**}$ & 0.5322$^{***}$ & -0.6331$^{***}$ & -0.4738$^{***}$ & 1 & ~ & ~ & ~ & ~ & ~ & ~ & ~ & ~ & ~ \\ 
			PG &	0.0924$^{*}$ & 0.5644$^{***}$ & -0.2808$^{***}$ & -0.1528$^{**}$ & 0.7504$^{***}$ & 1 & ~ & ~ & ~ & ~ & ~ & ~ & ~ & ~ \\ 
			JNJ &	0.4573$^{***}$ & 0.5669$^{***}$ & -0.4379$^{***}$ & -0.2835$^{***}$ & 0.7824$^{***}$ & 0.4341$^{***}$ & 1 & ~ & ~ & ~ & ~ & ~ & ~ & ~ \\ 
			OIL &	0.3493$^{***}$ & 0.5853$^{***}$ & -0.6032$^{***}$ & -0.5135$^{***}$ & 0.8831$^{***}$ & 0.5302$^{***}$ & 0.8046$^{***}$ & 1 & ~ & ~ & ~ & ~ & ~ & ~ \\ 
			WHEAT &	0.2084$^{***}$ & 0.5141$^{***}$ & -0.6787$^{***}$ & -0.5854$^{***}$ & 0.8468$^{***}$ & 0.5623$^{***}$ & 0.7137$^{***}$ & 0.9100$^{***}$ & 1 & ~ & ~ & ~ & ~ & ~ \\ 
			ZNM23 &	-0.2597$^{***}$ & -0.5172$^{***}$ & 0.7155$^{***}$ & 0.6348$^{***}$ & -0.8814$^{***}$ & -0.5346$^{***}$ & -0.7664$^{***}$ & -0.9465$^{***}$ & -0.8981$^{***}$ & 1 & ~ & ~ & ~ & ~ \\ 
			SILVER & 	0.012 & -0.2938$^{***}$ & 0.3056$^{***}$ & 0.4494$^{***}$ & -0.4883$^{***}$ & -0.5377$^{***}$ & -0.2296$^{***}$ & -0.4530$^{***}$ & -0.4421$^{***}$ & 0.5016$^{***}$ & 1 & ~ & ~ & ~ \\ 
			GOLD &	-0.5741$^{***}$ & -0.1756$^{***}$ & -0.3667$^{***}$ & -0.1677$^{***}$ & -0.1346$^{**}$ & 0.2643$^{***}$ & -0.1043$^{*}$ & 0.008 & 0.2110$^{***}$ & -0.033 & 0.2536$^{***}$ & 1 & ~ & ~ \\ 
			CCMP &	0.7984$^{***}$ & 0.7068$^{***}$ & 0.5825$^{***}$ & 0.6223$^{***}$ & 0.1693$^{***}$ & 0.2419$^{***}$ & 0.3897$^{***}$ & 0.2023$^{***}$ & 0.0303 & -0.041 & -0.007 & -0.4723 & 1 & ~ \\ 
			SX5E&	0.8005$^{***}$ & 0.6782$^{***}$ & 0.3206$^{***}$ & 0.3571$^{***}$ & 0.4124$^{***}$ & 0.3419$^{***}$ & 0.5725$^{***}$ & 0.4249$^{***}$ & 0.1995$^{***}$ & -0.2916$^{***}$ & -0.1215$^{**}$ & -0.5028$^{***}$ & 0.9041$^{***}$ & 1 \\ \hline
	\end{tabular}}
\vspace*{-0.3cm}	\caption{Correlation between BTC and traditional assets (20/8/2020 --  1/7/2022)} \label{correl-2-1st}
\end{table}

\vspace*{-0.4cm}
\renewcommand{\arraystretch}{1.3} 
\begin{table}[H]
	\fontsize{7pt}{7pt}\selectfont
	\centering
	\resizebox{\textwidth}{!}{
		\begin{tabular}{lllllllllllllll}
			\hline & \\[-1.5ex]
			&
			BTC & TSLA & NFLX & AMZN & KO & PG & JNJ & OIL & WHEAT & ZNM23 & SILVER & GOLD & CCMP & SX5E    \\ 	\hline	\hline & \\[-1.5ex]
			BTC &	1 & ~ & ~ & ~ & ~ & ~ & ~ & ~ & ~ & ~ & ~ & ~ & ~ & ~ \\ 
			TSLA & 	0.4625$^{***}$ & 1 & ~ & ~ & ~ & ~ & ~ & ~ & ~ & ~ & ~ & ~ & ~ & ~ \\ 
			NFLX &	-0.0057 & -0.6776$^{***}$ & 1 & ~ & ~ & ~ & ~ & ~ & ~ & ~ & ~ & ~ & ~ & ~ \\ 
			AMZN &	0.5396$^{***}$ & 0.9061$^{***}$ & -0.6243$^{***}$ & 1 & ~ & ~ & ~ & ~ & ~ & ~ & ~ & ~ & ~ & ~ \\ 
			KO &	0.0053 & 0.0187 & -0.0855 & 0.0868 & 1 & ~ & ~ & ~ & ~ & ~ & ~ & ~ & ~ & ~ \\ 
			PG &	-0.1615$^{*}$ & -0.3325$^{***}$ & 0.1678$^{*}$ & -0.2334$^{**}$ & 0.8862$^{***}$ & 1 & ~ & ~ & ~ & ~ & ~ & ~ & ~ & ~ \\ 
			JNJ &	-0.5115$^{***}$ & -0.3869$^{***}$ & -0.0964 & -0.3739$^{***}$ & 0.5586$^{***}$ & 0.6238$^{***}$ & 1 & ~ & ~ & ~ & ~ & ~ & ~ & ~ \\ 
			OIL &	0.3249$^{***}$ & 0.5943$^{***}$ & -0.7818$^{***}$ & 0.5898$^{***}$ & 0.1343 & -0.1611$^{*}$ & 0.0887 & 1 & ~ & ~ & ~ & ~ & ~ & ~ \\ 
			WHEAT &	-0.0784 & 0.4506$^{***}$ & -0.5450$^{***}$ & 0.3125$^{***}$ & -0.6074$^{***}$ & -0.7372$^{***}$ & -0.2562$^{**}$ & 0.4156$^{***}$ & 1 & ~ & ~ & ~ & ~ & ~ \\ 
			ZNM23 &	0.4490$^{***}$ & 0.4700$^{***}$ & -0.4555$^{***}$ & 0.5972$^{***}$ & 0.6617$^{***}$ & 0.4577$^{***}$ & 0.1813$^{*}$ & 0.5487$^{***}$ & -0.2809$^{***}$ & 1 & ~ & ~ & ~ & ~ \\ 
			SILVER & 	-0.1935$^{*}$ & -0.8158$^{***}$ & 0.7776$^{***}$ & -0.6900$^{***}$ & 0.2269$^{**}$ & 0.5357$^{***}$ & -0.2835$^{***}$ & -0.6520$^{***}$ & -0.6288$^{***}$ & -0.1823$^{*}$ & 1 & ~ & ~ & ~ \\ 
			GOLD &	0.20761$^{**}$ & -0.5669$^{***}$ & 0.6723$^{***}$ & -0.3722$^{***}$ & 0.3678$^{***}$ & 0.5710$^{***}$ & 0.0768 & -0.4172$^{***}$ & -0.7638$^{***}$ & 0.2084$^{**}$ & 0.8354$^{***}$ & 1 & ~ & ~ \\ 
			CCMP &	0.6902$^{***}$ & 0.6555$^{***}$ & -0.1618$^{*}$ & 0.7480$^{***}$ & 0.4621$^{***}$ & 0.2024$^{*}$ & -0.2785$^{***}$ & 0.3319$^{***}$ & -0.2431$^{**}$ & 0.7113$^{***}$ & -0.2581$^{**}$ & 0.1508 & 1 & ~ \\ 
			SX5E&	0.1553 & -0.5850$^{***}$ & 0.8704$^{***}$ & -0.5044$^{***}$ & 0.2686$^{***}$ & 0.4704$^{***}$ & 0.0049 & -0.6055$^{***}$ & -0.7780$^{***}$ & -0.0851 & 0.7946$^{**}$ & 0.8497$^{***}$ & 0.1278 & 1 \\ \hline
	\end{tabular}}
\vspace*{-0.3cm}	\caption{Correlation between BTC and traditional assets (01/7/2022 -- 24/2/2023)} \label{correl-2-2nd}
\end{table}

\end{document}